\documentclass[12pt]{article}

\usepackage{newtxtext,newtxmath}

\usepackage{graphicx}

\usepackage[letterpaper,margin=1in]{geometry}

\renewenvironment{abstract}
	{\quotation}
	{\endquotation}

\date{}

\makeatletter
\renewcommand{\fnum@figure}{\textbf{Figure \thefigure}}
\renewcommand{\fnum@table}{\textbf{Table \thetable}}
\makeatother

\usepackage{scicite}

\usepackage{url}

\newcommand{\CIV}{C\,{\small IV}\,}
\newcommand{\SiIV}{Si\,{\small IV}\,}
\newcommand{\NV}{N\,{\small V}\,}
\newcommand{\OVI}{O\,{\small VI}\,}
\newcommand{\HI}{H\,{\small I}\,}

\newcommand{\lb}{\lambda_{\mathrm{b}}}
\newcommand{\lr}{\lambda_{\mathrm{r}}}
\newcommand{\lu}{\lambda_{\mathrm{u}}}
\newcommand{\LL}{\lambda_{\mathrm{\ell}}}
\newcommand{\lm}{\lambda_{\mathrm{min}}}
\newcommand{\lv}{\lambda_{\mathrm{v}}}
\newcommand{\lbol}{\ifmmode L_{\rm bol}\else $L_{\rm bol}$\fi}
\newcommand{\kms}{\ifmmode\mathrm{\,km\,s^{-1}}\else\,km\,s$^{-1}$\fi}
\newcommand{\cm}{{\,{\rm cm}}}
\newcommand{\g}{{\,{\rm g}}}

\newcommand{\ergs}{\ifmmode\mathrm{\,erg\,s^{-1}}\else\,erg\,s$^{-1}$\fi}
\newcommand{\sunit}{{{\rm s}}}
\newcommand{\Lkin}{\ifmmode L_{\text{kin}}\else $L_{\text{kin}}$\fi}
\newcommand{\vw}{\ifmmode v_\infty\else $v_\infty$\fi}

\def\scititle{
	Radiation pressure powers quasar broad absorption line winds but fails to drive galaxy feedback 
}
\title{\bfseries \boldmath \scititle}

\author{
	B.~Seefeldt-Gail$^{1,2,3\ast}$,
	L.~Mas-Ribas$^{4}$,
	N.~Murray$^{1}$\and
	\small$^{1}$Canadian Institute for Theoretical Astrophysics, University of Toronto, Toronto, ON, Canada.\and
	\small$^{2}$David A. Dunlap Department of Astronomy and Astrophysics, University of Toronto, Toronto, ON, Canada.\and
	\small$^{3}$Dunlap Institute for Astronomy and Astrophysics, University of Toronto, Toronto, ON, Canada.\and
	\small$^{4}$Department of Astronomy and Astrophysics, University of California Santa Cruz, Santa Cruz, CA, USA.\and
	\small$^\ast$Corresponding author email: braden.gail@mail.utoronto.ca
}

\begin{document} 

\maketitle

\begin{abstract} \bfseries \boldmath
Most cosmological simulations invoke quasar-driven winds as the primary mechanism quenching star formation in massive galaxies, assuming wind momentum flux approaches the radiation-pressure maximum, $L/c$. Using 81,388 broad absorption line troughs from the SDSS-IV DR16 quasar catalog and detailed mock trough spectra, we detect line locking, a spectral signature of radiation-pressure-driven outflows, in 60 to 70 percent of troughs, demonstrating that radiative line driving dominates wind acceleration and ruling out alternative mechanisms. Direct measurement of the wind energy budget yields a median kinetic luminosity of only 0.01 percent of bolometric luminosity, 50 times below the level required for effective feedback. Line-driven quasar winds are therefore insufficiently powerful to quench star formation, challenging the primary physical mechanism invoked for black hole feedback in galaxy formation models.
\end{abstract}

\noindent
The ``overcooling problem,'' in which early galaxy formation models without feedback mechanisms predicted far too many massive, actively star-forming galaxies, motivated the adoption of active galactic nucleus (AGN) feedback as a cornerstone of modern cosmological simulations \cite{1978MNRAS.183..341W,1991ApJ...367...45C,1991ApJ...379...52W}. Over the last decade, simulations modeling galaxy populations across cosmic time and over cosmological volumes have incorporated AGN feedback prescriptions to reproduce the observed scaling relations between black hole mass and host galaxy properties \cite{2007MNRAS.380..877S,2014MNRAS.444.1518V,2015MNRAS.450.1937C}. In practice, ``quasar mode'' feedback in these simulations is understood to operate through broad absorption line (BAL) winds -- high velocity, blueshifted outflows observed in 20 to 40 percent of luminous quasars \cite{2023ApJ...952...44B} -- with sub-grid feedback prescriptions tuned to match observed black hole and galaxy scaling relations under the assumption that wind momentum fluxes approach $\lbol/c$ \cite{2010MNRAS.401....7H,2014MNRAS.438.3024L}. These prescriptions are calibrated to reproduce observations rather than derived from first principles of wind physics, and whether the assumed wind properties are physically achievable has remained an open question.

Several physical mechanisms have been proposed to drive BAL winds, including radiation pressure on resonance lines, radiation pressure on electrons, or on dust grains, magnetic fields, and thermally driven winds \cite{1998A&A...331L...1S,Murray1995,2003ApJ...596L..27K,Netzer2007,1982MNRAS.199..883B}. Identifying which mechanism dominates has important implications, as the dominant mechanism sets the maximum momentum and energy the wind can deliver to its host galaxy, and therefore determines whether quasar winds can plausibly produce the feedback invoked in galaxy formation simulations.

Line locking provides a decisive observational test of the driving mechanism. When two gas clouds in a radially accelerating outflow are separated by a relative velocity equal to the rest-frame wavelength splitting of a resonance doublet, e.g., the $497\kms$ splitting of the \CIV\ doublet, the upstream cloud casts a shadow on the downstream cloud at the wavelength of its red doublet component. This shadow reduces the downstream cloud's radiative acceleration and can lock the relative velocity of the two clouds at the doublet splitting velocity \cite{Braun1989,Mas-Ribas2019}. This phenomenon arises uniquely through radiation pressure acting on resonance lines, and as such, its detection in an outflow constitutes direct evidence that line driving dominates the acceleration of that outflow, simultaneously constraining the contribution of any other proposed mechanism.

Earlier work \cite{Mas-Ribas2019} reported a ubiquitous line-locking signature in stacked BAL spectra, but the physical reality of the detection could not be confirmed because stacking artifacts, such as contamination from accidentally aligning on the red doublet component rather than the blue, could in principle mimic the signal. Here we present a definitive detection of \CIV\ line locking using 81,388 BAL troughs drawn from the SDSS-IV DR16 quasar catalog \cite{DR16}, measuring a line locking fraction of 60 to 70 percent and validating the result against an extensive kernel-density-estimation-based mock spectral framework. We then use the confirmed dominance of radiative line driving to derive a robust measurement of the kinetic energy budget of these winds. Our results directly challenge the physical basis of quasar-mode AGN feedback as implemented in many current galaxy formation models.

\subsection*{Most broad absorption line outflows show line locking}

\begin{figure} 
	\centering
	\includegraphics[width=0.62\textwidth]{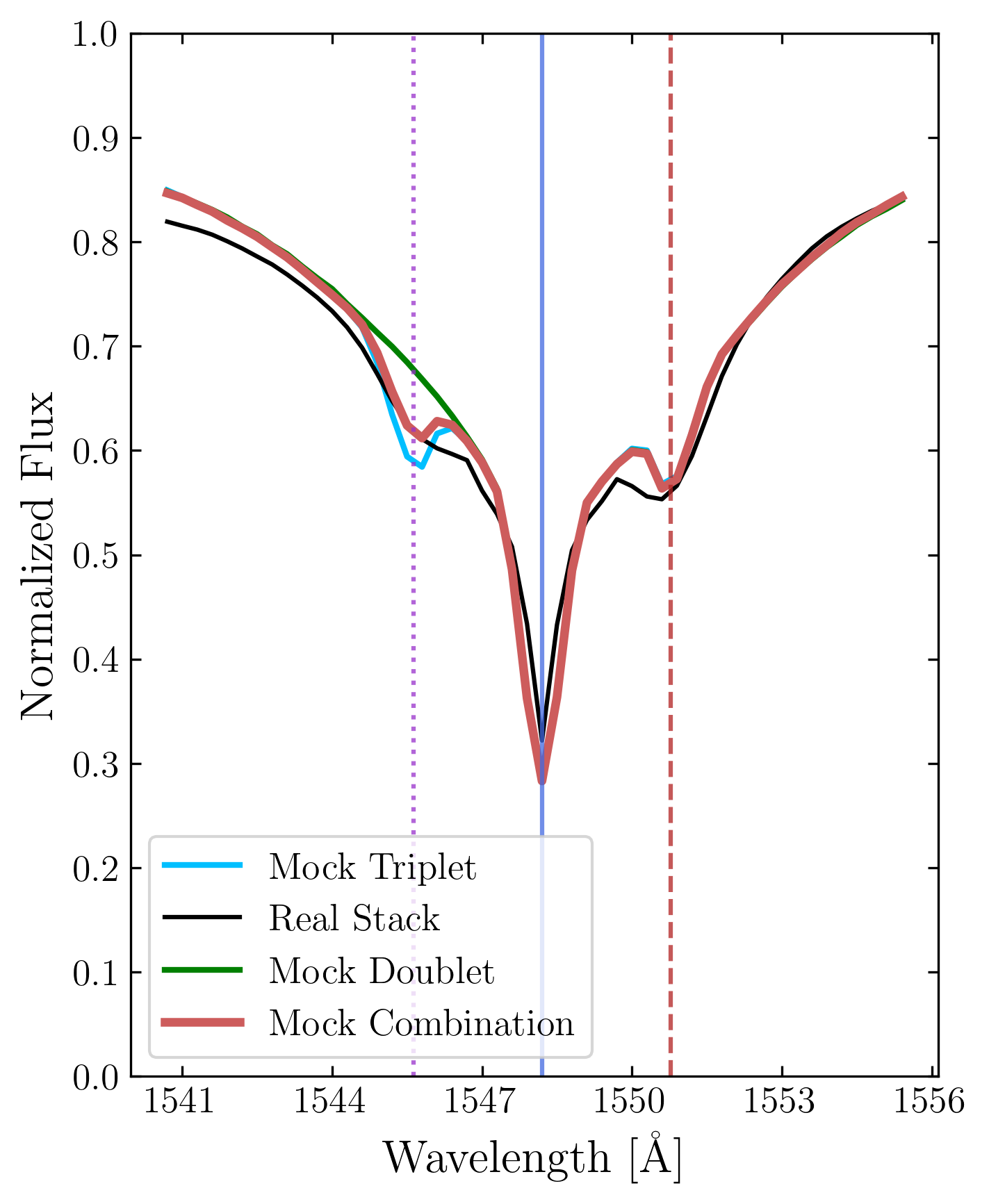}
	\caption{\textbf{Detection of \CIV\ line locking in stacked broad absorption line quasar spectra.}
Weighted composite spectra constructed from 29,332 absorption troughs, each centered at
its flux minimum. The black line shows the real SDSS DR16 data stack, the green line shows
mocks without line locking, the light blue line shows mocks in which 100 percent of troughs
are line-locked, and the red line shows the best-fit combination of two thirds line-locked
and one third non-line-locked mocks. Vertical lines mark the \CIV\ blue doublet wavelength
$\lb$ (solid blue, 1548.20 \AA), red doublet wavelength $\lr$ (dashed red, 1550.77 \AA),
and the expected line-locking wavelength $\lv$ (dotted purple, 1545.63 \AA). The doublet-only
mock (green) fails to reproduce the absorption at $\lv$ visible in the real data, while the
combined mock (red) matches it closely, demonstrating that the $\lv$ feature arises from
genuine line locking rather than a stacking artifact.}
	\label{fig:detection}
\end{figure}

We searched for \CIV\ ($\lambda\lambda$1548, 1551 \AA) broad ($>450$ \kms) absorption troughs in the Sloan Digital Sky Survey (SDSS-IV) Data Release 16 (DR16) quasar catalog \cite{DR16}, which contains approximately 750,000 confirmed quasars spanning redshifts $0 \lesssim z \lesssim 5$. Using the \texttt{baltools} software \cite{Convolution_Neural_Network_PCA,Martini2025} to identify and characterize absorption troughs, and applying quality cuts on continuum fit quality, signal-to-noise ratio, and BAL probability (Table~\ref{tab:selection}; see Materials and Methods), we assembled a final sample of 81,388 \CIV\ absorption troughs. Constructing weighted composite (stacked) spectra by centering each trough on its flux minimum (see Materials and Methods), we detect a triplet absorption feature with components at the \CIV\ blue doublet wavelength ($\lb=1548.20$ \AA), red doublet wavelength ($\lr=1550.77$ \AA), and the wavelength expected for line-locked absorption ($\lv=1545.63$ \AA, corresponding to a velocity separation of 497 \kms; Fig.~\ref{fig:detection}).

To rigorously verify that the observed signal at $\lv$ is physical rather than a stacking artifact, we constructed 29,332 mock absorption spectra without line locking using a six-parameter Gaussian kernel density estimator (KDE) trained on the real trough parameter distributions, including their pairwise correlations (see Materials and Methods). When stacked, these mocks reproduce the \CIV\ doublet structure but produce no absorption at $\lv$ (Fig.~\ref{fig:detection}, green line). Three independent tests confirm the physical origin of the signal (see Materials and Methods): accidental red-doublet-line contamination in the stack is below 1 percent; the feature persists independent of trough width, ruling out extended trough morphology; and mocks incorporating a line-locked third absorption component reproduce the real data only when that component is present.

\begin{figure}
	\centering
	\includegraphics[width=0.8\textwidth]{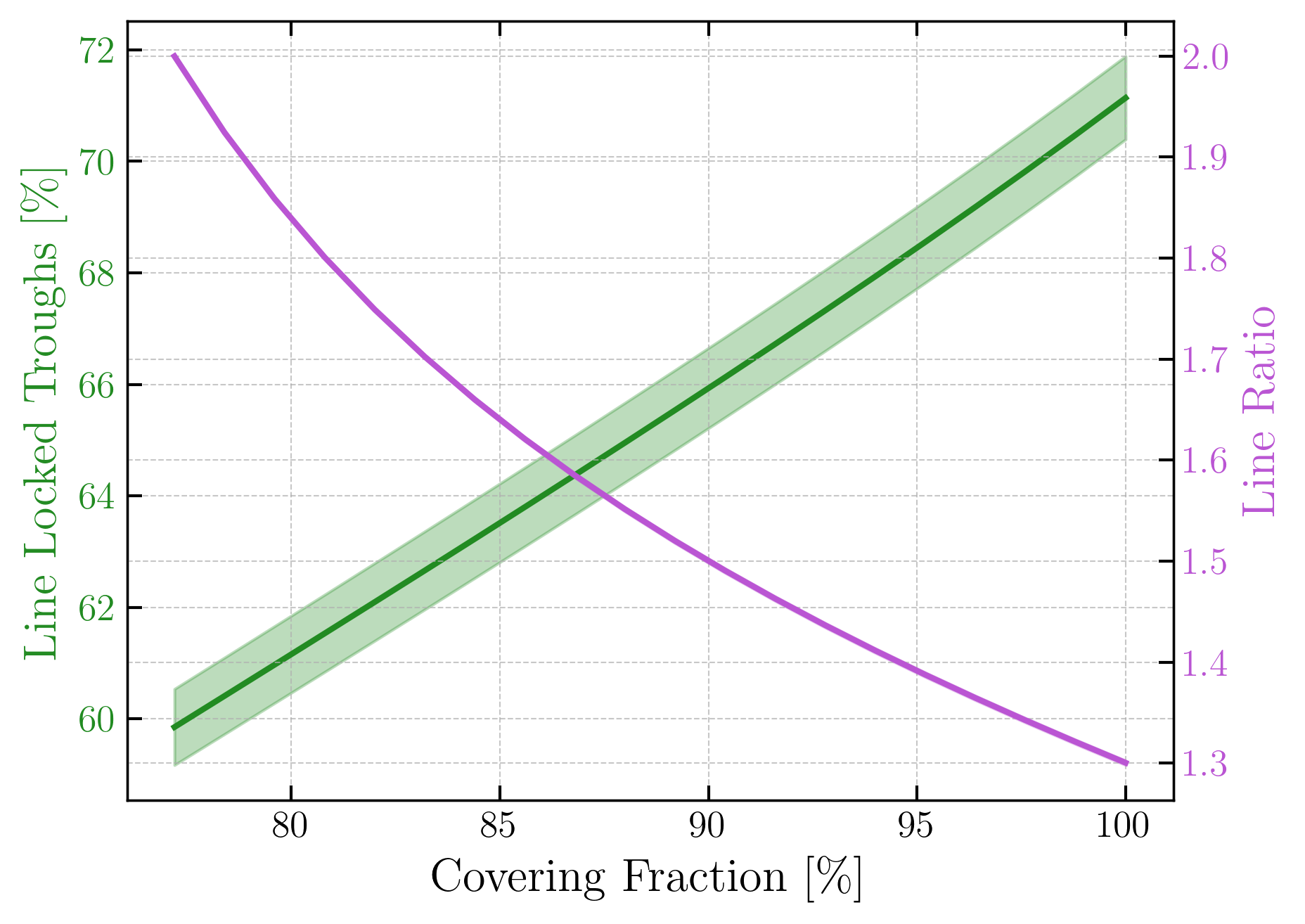}
	\caption{\textbf{Line locking fraction and \CIV\ doublet optical depth ratio as functions of covering fraction.}
Results from fitting a physically motivated three-component model, combining line-locked
two-cloud systems, single-cloud systems, and misalignment contamination, to the real stacked
spectrum evaluated at $\lb$, $\lr$, and $\lv$. The green line and shading show the fraction
of troughs exhibiting line locking (left axis), while the violet line shows the \CIV\ doublet
optical depth ratio $\tau_b/\tau_r$ (right axis). Line widths correspond to contamination
fractions between 0 and 1 percent and we note that the error band on the doublet ratio is narrower than the
plotted line width and is not visible. The line locking fraction is robustly 60 to 70 percent
across the full range of covering fractions and contamination levels explored.}
	\label{fig:fraction}
\end{figure}

To quantify the prevalence of line locking, we fit the stacked spectrum with a physically motivated model combining line-locked two-cloud systems, single-cloud systems, and contamination from misaligned stacking (see Materials and Methods). Evaluating this model at $\lb$, $\lr$, and $\lv$ for covering fractions between 80 and 100 percent and contamination fractions between 0 and 1 percent, we measure a line locking fraction of 60 to 70 percent (Fig.~\ref{fig:fraction}). The fitted \CIV\ optical depths are of order unity, consistent with theoretical expectations for line-driven winds at high outflow velocities \cite{Murray1995}. This prevalence, approximately two thirds of all BAL troughs exhibiting a sustained velocity lock, indicates that radiative line driving is the dominant acceleration mechanism for the majority of quasar outflows. Combined with the 20 to 40 percent of luminous quasars that exhibit broad absorption troughs, this implies that radiative line driving observably operates in roughly one quarter of all quasars, a lower limit given the factors that can suppress the line locking signature even when present (Supplementary Text).

\subsection*{Broad absorption line winds cannot power galaxy feedback}

Having established that radiative line driving dominates the acceleration of these outflows, we measured their momentum and energy budget. For 116,428 troughs identified in 73,442 quasars, we measured the absorbed \CIV\ flux by integrating the area between the quasar continuum and the observed trough spectrum, and extended this measurement across the full rest-frame range of 900 to 1600 \AA\ to account for absorption from \NV, \OVI, and other ionic species, using velocity-dependent corrections for the fractional contribution of \CIV\ to the total absorbed flux (see Materials and Methods). Applying a covering fraction $C_f=0.38$, derived from the fraction of DR16 quasars exhibiting outflow absorption features, and invoking momentum conservation under the assumption of single scattering, we derived a mass outflow rate for each trough.

\begin{figure}
	\centering
	\includegraphics[width=0.95\textwidth]{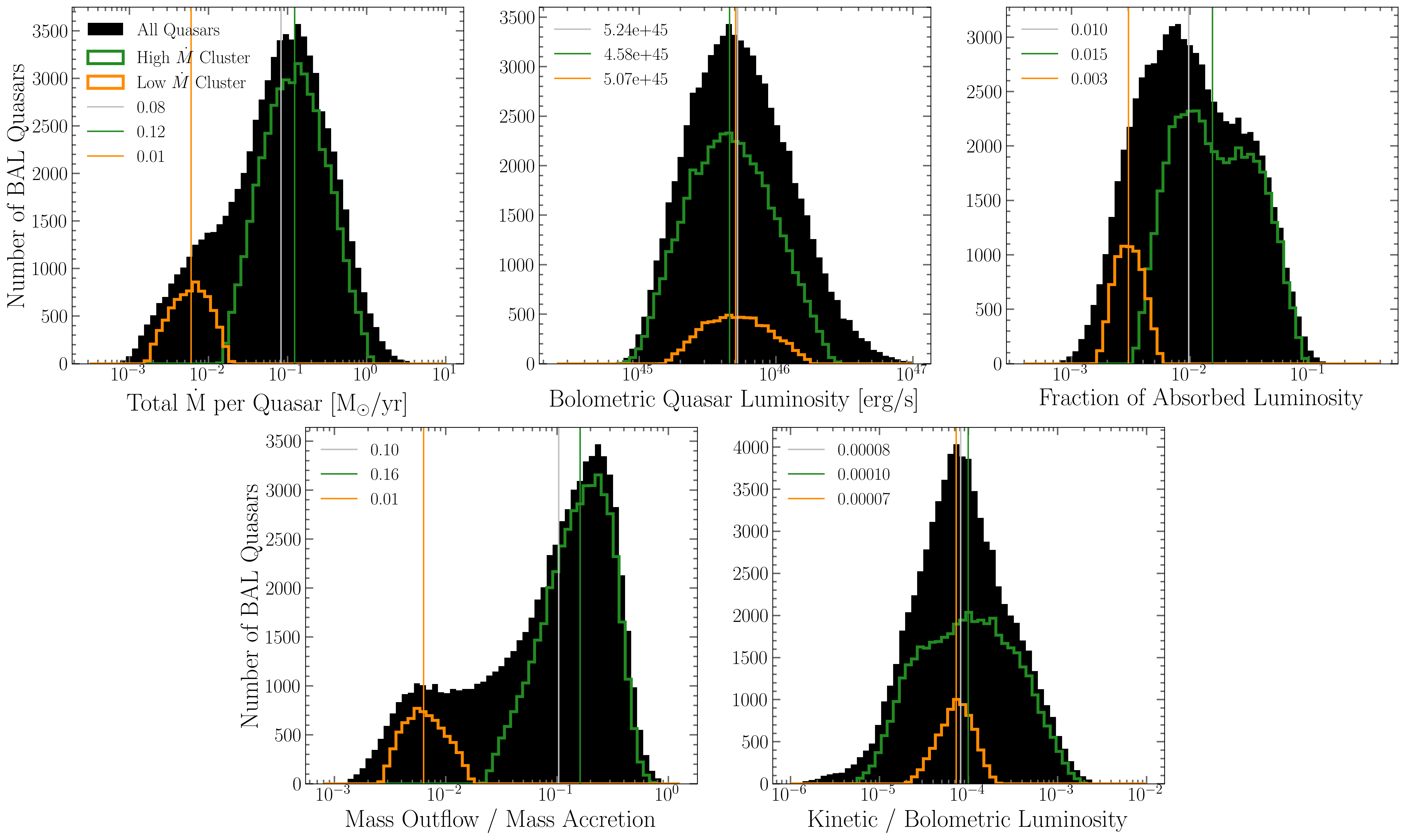}
	\caption{\textbf{Distributions of mass outflow rate, luminosity, and kinetic energy for the quasar sample.}
Each panel shows the full sample distribution (black) together with the two populations
identified by density-based clustering: a high-mass-outflow-rate population (green) and
a low-mass-outflow-rate population (orange). Vertical lines mark median values. From left
to right and top to bottom: total mass outflow rate per quasar (median 0.08 $M_\odot$
yr$^{-1}$); quasar's bolometric luminosity; fraction of bolometric luminosity absorbed by
the outflow (median approximately 1 percent); ratio of mass outflow rate to mass accretion
rate (median approximately 12 percent); and ratio of wind kinetic luminosity to bolometric
luminosity (median $10^{-4}$, or 0.01 percent), which falls 50 to 500 times below the
0.5 to 5 percent range required for effective AGN feedback. All mass outflow rates and
kinetic luminosities assume the single scattering limit.} 
	\label{fig:histograms}
\end{figure}

The resulting distribution of mass outflow rates spans $\sim10^{-3}$ to 5 $M_\odot$ yr$^{-1}$, with a median of 0.08 $M_\odot$ yr$^{-1}$, corresponding to approximately 10 percent of the inferred mass accretion rate (Fig.~\ref{fig:histograms}). The absorbed luminosity comprises approximately 1 percent of the bolometric luminosity, and the resulting wind kinetic luminosities, $\Lkin = \frac{1}{2}\dot M v^2$, have a median value of $\sim$0.01 percent of $\lbol$ (Fig.~\ref{fig:histograms}).

Current theoretical models and cosmological simulations of AGN feedback require kinetic luminosities of 0.5 to 5 percent of $\lbol$ to reproduce the observed relation between black hole mass and host galaxy velocity dispersion, and to quench star formation in massive galaxies \cite{2010MNRAS.401....7H,2014MNRAS.438.3024L}. Our measured median value falls 50 to 500 times below this required range. This shortfall persists under the most generous possible correction: assuming 100 percent covering fraction rather than 38 percent raises $\Lkin/\lbol$ to only 0.03 percent, still an order of magnitude below the lowest feedback threshold invoked in the literature.

The demonstrated dominance of radiative line driving is what makes this shortfall a firm conclusion rather than a lower limit subject to revision. If the dominant driving mechanism were uncertain, a low measured momentum flux could in principle be rescued by invoking unmeasured contributions from dust radiation pressure, magnetic driving, or thermal pressure. The 60 to 70 percent line locking fraction instead demonstrates observationally that radiation pressure on lines is the dominant mechanism operating in the large majority of these outflows, which places a firm upper limit on the contribution any alternative mechanism can be making. A kinetic luminosity deficit of 50 to 500 times cannot be closed by appealing to mechanisms that the data show are not operating at significant levels.

Density-based clustering of the outflow sample (DBSCAN \cite{Ester1996}; see Materials and Methods) reveals two distinct populations (Fig.~\ref{fig:histograms}). A high-mass-outflow-rate population exhibits wider, slower troughs, a clear line locking signature, and significant dust reddening. A low-mass-outflow-rate population consists of narrow ($<$2,000 \kms), high-velocity ($>$8,000 \kms) troughs with no detectable line locking and minimal dust extinction, properties more consistent with intervening absorbers than with genuine outflowing wind material (Supplementary Text). The kinetic luminosity deficit is concentrated in the high-mass population where line locking is confirmed, demonstrating that our feedback conclusion applies specifically to genuine line-driven winds.

\subsection*{Radiative driving dominates the acceleration region}

\begin{figure}
	\centering
	\includegraphics[width=0.75\textwidth]{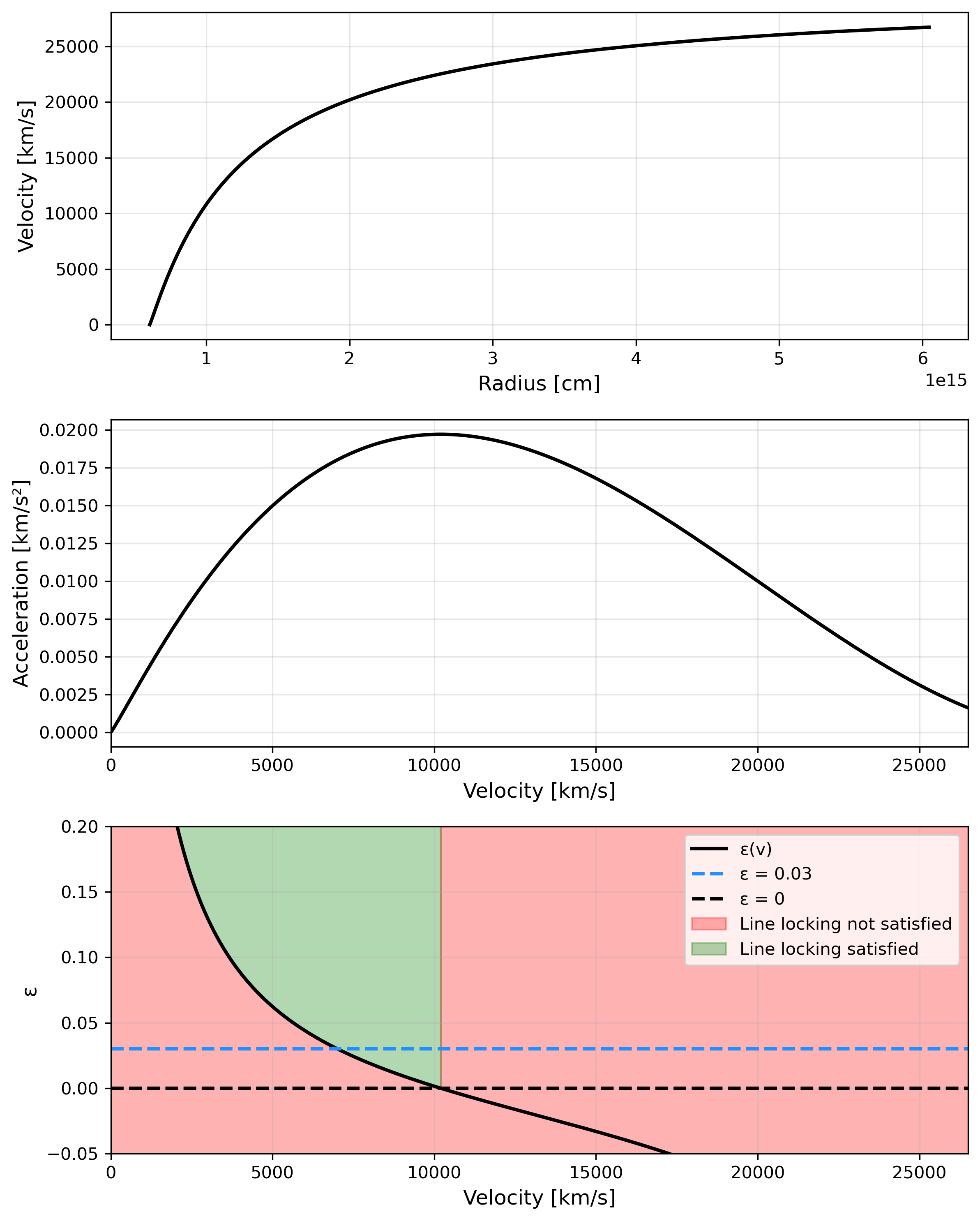}
	\caption{\textbf{CAK line-driven wind dynamics and the line locking condition.}
All panels show results from the standard Castor, Abbott \& Klein (CAK)
line-driven wind model \cite{CAK1975,Murray1995} for fiducial quasar parameters
$M_\bullet=10^8\,M_\odot$, $\lbol=10^{46}\,\ergs$, $\alpha=0.8$, $\xi=10^{-3}$
(see Materials and Methods). Top: wind velocity as a function of launch radius.
Middle: wind acceleration as a function of outflow velocity. Bottom: the fractional
acceleration difference between two gas clouds separated by the \CIV\ doublet
velocity, as a function of outflow velocity. Green and red regions indicate where
the line locking condition is and is not satisfied, respectively. The dashed blue
line shows our estimated $\epsilon\approx0.03$, which permits active line locking
up to approximately 10,000 \kms.}
	\label{fig:dynamics}
\end{figure}

To characterize the physical conditions required for line locking, we modeled the wind kinematics using the standard Castor, Abbott \& Klein (CAK) formalism for line-driven winds \cite{CAK1975}, adapted for the geometry and rotational support of an AGN accretion disk \cite{Murray1995} (see Materials and Methods). For fiducial parameters appropriate to a luminous quasar ($M_\bullet=10^8\,M_\odot$, $\lbol=10^{46}\,\ergs$), the wind is launched at a radius $r_0\approx10^{15}$ cm and reaches a terminal velocity of order 30,000 \kms\ (Fig.~\ref{fig:dynamics}).

Line locking requires that the fractional reduction in acceleration produced by shadowing, $\epsilon$, exceed the relative difference in acceleration between two clouds separated by the \CIV\ doublet velocity \cite{Braun1989}. We estimate $\epsilon$ from three independently measurable quantities (see Materials and Methods): the relative importance of line driving compared to other forces ($g_\ell/g_{\rm other}\sim1$, justified self-consistently below); the fractional contribution of \CIV\ to the total absorbed flux ($F_{\rm CIV}/F_{\rm other}\approx0.15$, measured directly from our extended absorption spectra); and the fraction of \CIV\ photon flux actually blocked during a lock ($F_{\rm masked}\approx0.2$, from the measured difference between line-locked and single-cloud absorption profiles). Together these give $\epsilon\approx0.03$, which satisfies the line locking condition for outflow velocities greater than $\approx 7,000\kms$ up to approximately 10,000 \kms\ (Fig.~\ref{fig:dynamics}, green region).

\begin{figure}
	\centering
	\includegraphics[width=0.95\textwidth]{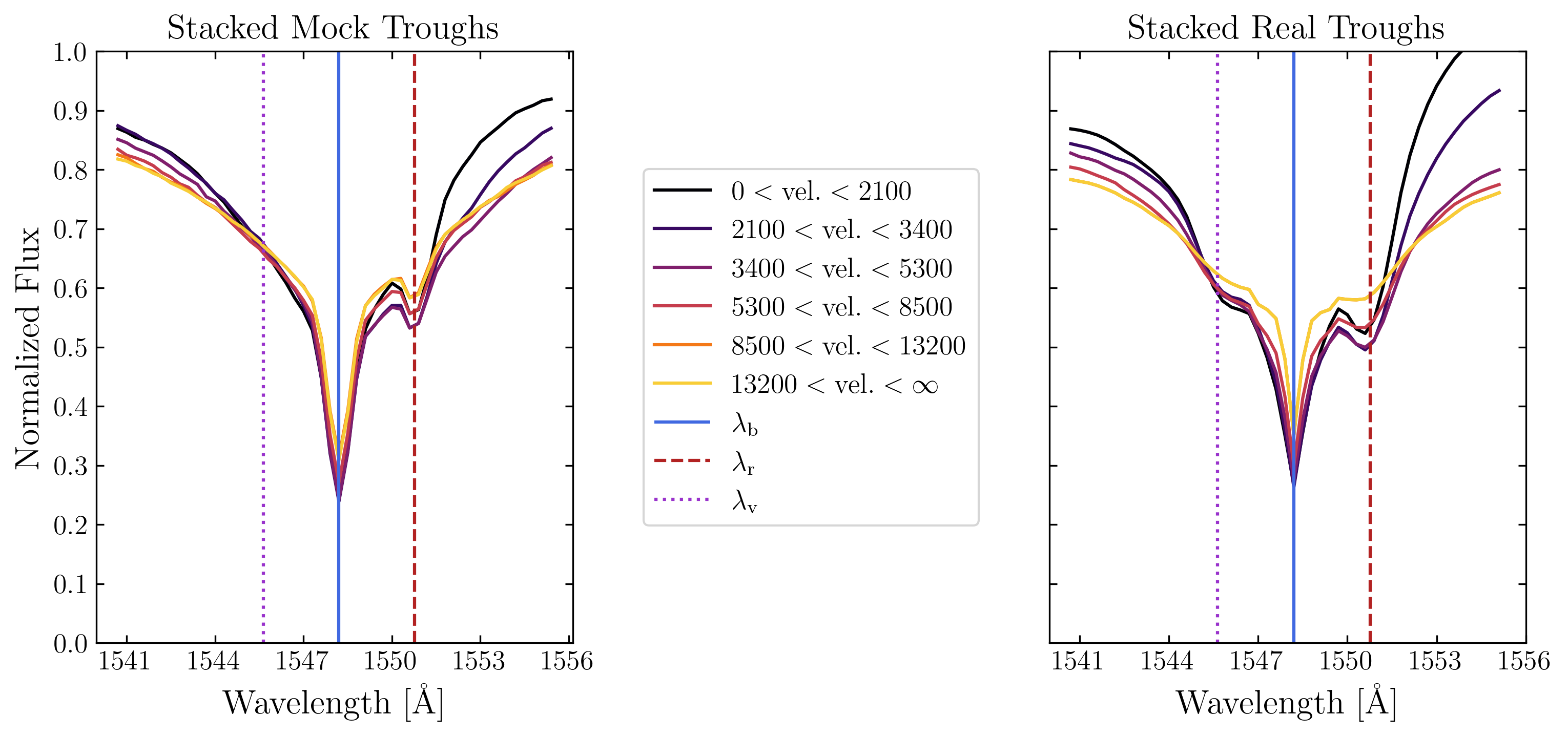}
	\caption{\textbf{Line locking persists across a wide range of outflow velocities.}
Stacked \CIV\ absorption troughs in six outflow velocity bins for mock spectra
without line locking (left) and real SDSS DR16 data (right), each containing
approximately 3,900 spectra. Vertical lines mark $\lb$ (solid blue), $\lr$
(dashed red), and $\lv$ (dotted purple). An absorption feature at $\lv$ is
visible across the full velocity range in the real data, from below 2,100 \kms\
to above 13,200 \kms, while no such feature appears in any mock stack.}
	\label{fig:velocity_stacks}
\end{figure}

We observe line locking signatures across a substantially wider range, from below 2,100 \kms\ to above 13,200 \kms\ (Fig.~\ref{fig:velocity_stacks}). This broad observed velocity range is inconsistent with any scenario in which line driving is subdominant: if continuum scattering, magnetic forces, or dust absorption contributed comparably to the wind acceleration, $\epsilon$ would be substantially smaller, restricting line locking to a much narrower velocity range than is observed. The wide velocity range over which line locking is detected therefore self-consistently validates the assumption $g_\ell/g_{\rm other}\sim1$ on which our mass outflow measurement depends.

The persistence of the signature beyond $\sim$10,000 \kms\ is consistent with finite cloud dispersion timescales: gas clouds leaving the actively line-locked region at $\sim$10,000 \kms\ continue to drift with their relative velocity separation intact, and can travel to $\sim$22,000 \kms\ before the configuration disperses (see Materials and Methods).
The observation of line locking at velocities below $7,000\kms$ shows that our simple dynamical model is inadequate. One clear oversimplification that might be relevant is the assumption of purely radial flow lines. Since the likely source of the outflowing gas is the accretion disk, we would expect the flow to be non-radial, particularly near the launching radii. In that case, the line of sight velocity may be well below the local wind speed, resulting in line locking appearing to occur at very low velocity.

\subsection*{A challenge to galaxy formation models}

Our results present a direct empirical challenge to a foundational assumption of current galaxy
formation models. Simulations reproduce the observed relations between black hole mass and
host galaxy properties by invoking AGN feedback winds carrying momentum flux approaching
$\lbol/c$ \cite{2007MNRAS.380..877S,2014MNRAS.444.1518V,2015MNRAS.450.1937C,2010MNRAS.401....7H,2014MNRAS.438.3024L}. We show that the kinetic luminosity of line-driven quasar winds,
the mechanism conventionally invoked for this feedback, falls 50 to 500 times below the level
simulations require to reproduce observed black hole and galaxy scaling relations. Reproducing
the observed feedback energy budget therefore requires either a different physical mechanism
or a substantial revision to how feedback is implemented.

Two physical mechanisms remain plausible candidates for supplying the additional feedback energy galaxy formation models require. Ultra-fast outflows detected in X-ray spectra at velocities of order one tenth the speed of light \cite{2010A&A...521A..57T, Matzeu2023,Gianolli2014} potentially carry substantially more kinetic energy than the UV-absorbing BAL winds characterized here. Radio jets are established drivers of feedback in galaxy groups and clusters \cite{2007ARA&A..45..117M} and may contribute during the quasar phase in radio-loud systems. Our results shift the observational burden onto these mechanisms, and motivate both comprehensive characterization of ultra-fast outflow energetics and theoretical work connecting small-scale jets to galaxy-scale quenching.

Line-driven BAL winds remain a ubiquitous and now physically well-characterized component of quasar activity. Direct observational proof of their dominant driving mechanism, combined with a measurement of their resulting energy budget, demonstrates that they cannot supply the feedback energy current cosmological simulations require, fundamentally constraining the landscape of viable AGN feedback physics in galaxy formation.


\clearpage 

%
\bibliography{line_locking}
\bibliographystyle{sciencemag}



\section*{Acknowledgments}
We are grateful to Paul Martini and Allyson Brodzeller for their help with \textsc{baltools}, for sharing their PCA templates, and for discussions about BALs and quasar continua. We also thank Martin Gaskill and Xavier Prochaska for their comments and discussions, and Gwendolyn Eadie for her insight into the statistical analysis and comparison of our mock spectra. We thank Alicia Savelli for insightful conversations throughout the project and for her careful reading of the manuscript, and Matthew Nesom for careful proofreading.
\paragraph*{Funding:}
B. Seefeldt-Gail and N. Murray acknowledge the support of the Natural Sciences and Engineering Research Council of Canada (NSERC), RGPIN-2023-04901.
\paragraph*{Author contributions:}
B.S.-G. led the data analysis, developed the KDE-based mock spectral framework, performed the line locking detection and statistical characterization, and carried out the mass outflow rate calculations. L.M.-R. contributed to the spectroscopic analysis methodology and outflow property measurements. N.M. contributed to the theoretical framework for line-driven wind dynamics and the physical interpretation of the results. All authors contributed to writing the manuscript.
\paragraph*{Competing interests:}
There are no competing interests to declare.
\paragraph*{Use of AI-assisted technology:}
A large language model (Claude, Anthropic) was used an aid in editing prose, and restructuring and reformatting the text for this journal. It was not used to generate or analyze data, produce figures, or draw scientific conclusions. The authors take full responsibility for the content of this work.
\paragraph*{Data, code and materials availability:}
SDSS-IV DR16 quasar spectra are publicly available at \url{https://www.sdss.org/dr16/}. The \texttt{baltools} software used to identify absorption troughs is publicly available at \url{https://github.com/paulmartini/baltools}. Analysis code for the mock spectral construction, KDE framework, and stacking
pipeline is available at \url{https://github.com/BradenSeefeldt/Line-Locking-Final-Code} and archived at
\url{https://doi.org/10.5281/zenodo.21417626} \cite{zenodo_code}.


\subsection*{Supplementary materials}
Materials and Methods\\
Supplementary Text\\
Figs. S1 to S17\\
Tables S1 and S2\\
References \textit{(24-\arabic{enumiv})}


\newpage


\renewcommand{\thefigure}{S\arabic{figure}}
\renewcommand{\thetable}{S\arabic{table}}
\renewcommand{\theequation}{S\arabic{equation}}
\renewcommand{\thepage}{S\arabic{page}}
\setcounter{figure}{0}
\setcounter{table}{0}
\setcounter{equation}{0}
\setcounter{page}{1} 


\begin{center}
\section*{Supplementary Materials for\\ \scititle}

B.~Seefeldt-Gail$^{\ast}$,
L.~Mas-Ribas,
N.~Murray\\
\small$^\ast$Corresponding author. Email: braden.gail@mail.utoronto.ca
\end{center}

\subsubsection*{This PDF file includes:}
Materials and Methods\\
Supplementary Text\\
Figures S1 to S17\\
Tables S1 and S2

\newpage


\subsection*{Materials and Methods}

\subsubsection*{Line locking formalism}

We focus on line locking caused by an atomic doublet, where two spectral lines arise from transitions between a common ground state and two excited states differing only in total angular momentum. Relevant doublets in AGN outflows include \CIV, \SiIV, \OVI, and \NV\ \cite{Braun1989,Chen2025}.

\begin{figure}
	\centering
	\includegraphics[width=0.45\textwidth]{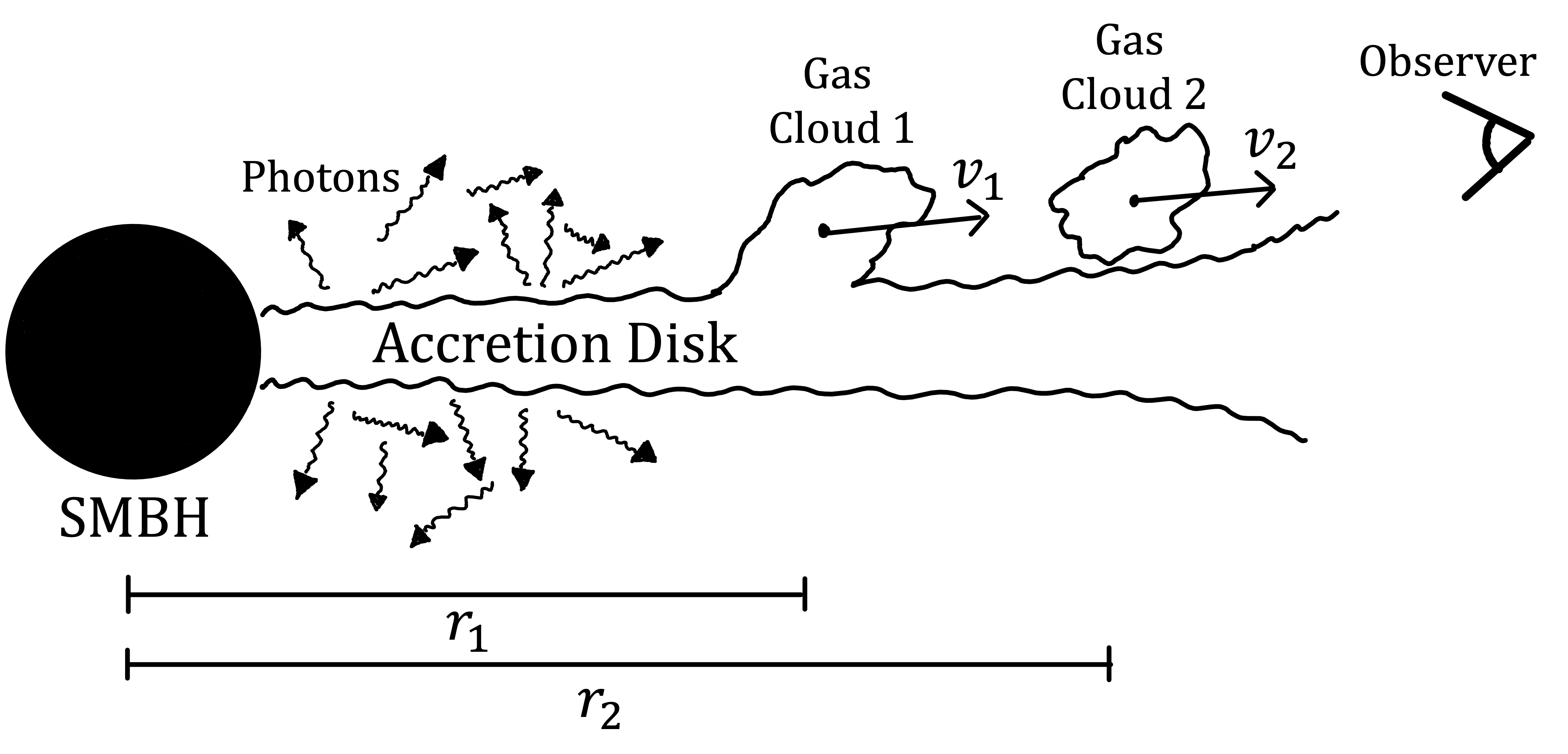}
	\caption{\textbf{Schematic of the two-cloud line locking geometry.}
	Two gas elements travel along the line of sight away from a supermassive black hole. Radiation from the inner accretion disk accelerates the two clouds, situated at radii $r_1$ and $r_2$, with velocities $v_1$ and $v_2$. \CIV\ line locking occurs when the relative velocity between the two clouds is approximately 497 \kms.}
	\label{fig:s_schematic}
\end{figure}

We consider two absorbing gas elements (``clouds''), at radii $r_1(t)$ and $r_2(t)>r_1(t)$, moving along the line of sight with velocities $v_1(t)$ and $v_2(t)>v_1(t)$ and accelerations $a_1(t)$ and $a_2(t)$, respectively (Fig.~\ref{fig:s_schematic}). For a given doublet we define the two transition wavelengths in the wind frame such that $\lr>\lb$. We are interested in the shadowing of the red wing of the doublet in cloud 2 by gas in cloud 1 scattering radiation via the blue doublet component. This shadowing occurs due to the Doppler shift corresponding to the velocity difference between the two clouds,
\begin{equation}
\lambda=\lambda_0\frac{\sqrt{1+\beta}}{\sqrt{1-\beta}}\quad,\quad\beta=\frac{v}{c}~,
\label{eq:doppler}
\end{equation}
where $\lambda$ corresponds to either $\lr$ (cloud 2) or $\lb$ (cloud 1), and $\lambda_0$ is the wavelength in the AGN frame.

For $r_1<r_2$ in a radially accelerating outflow, line locking can occur when the upstream (slower) gas at $r_1$ absorbs light at the wavelength corresponding to the red doublet component of the downstream (faster) gas at $r_2$. The downstream element then experiences a reduction in its acceleration $a_2$ by an amount $\epsilon a_2$ relative to an unshadowed parcel. If this reduction is large enough that $a_2=a_1$, the relative velocity between the two parcels vanishes, producing a metastable state in which both velocities increase with time while their relative velocity remains fixed.

The equilibrium velocity offset is
\begin{equation}
\Delta v_0=\pm c\,\frac{\lr^2-\lb^2}{\lr^2+\lb^2}~,
\label{eq:dv0}
\end{equation}
where the sign depends on whether $a_1<a_2$ or $a_1>a_2$. For the \CIV, \NV, \OVI, and \SiIV\ doublets, $\Delta v_0\approx$ 497, 962, 1632, and 1933 \kms, respectively.

The difference in acceleration between the two clouds at the onset of shadowing (when $\Delta v<\Delta v_0$) is $\Delta\dot v_{\rm before}=a_2-a_1$ \cite{Braun1989}, while at the point of maximal shadowing it is $\Delta\dot v_{\rm max}=(a_2-\epsilon_{\rm max}|a_2|)-a_1$. During an actual lock, $\epsilon<\epsilon_{\rm max}$, but is just large enough that $\Delta\dot v_{\rm lock}=0$. Line locking is therefore possible only if $\Delta\dot v_{\rm max}$ has the opposite sign to $\Delta\dot v_{\rm before}$, giving the condition
\begin{equation}
\text{if}\quad a_2>a_1,\quad\text{then}\quad\epsilon>\frac{a_2-a_1}{|a_2|}~.
\label{eq:llcondition}
\end{equation}
This condition is used below to determine the velocity range over which line locking can be sustained.

\subsubsection*{Quasar spectra and trough selection}

We use spectra from the SDSS-IV Data Release 16 (DR16) quasar catalog \cite{DR16}, containing approximately 750,000 confirmed quasars spanning $0\lesssim z\lesssim5$. DR16 of the extended Baryon Oscillation Spectroscopic Survey \cite{SDSSIV_Survey} improves on earlier releases \cite{DR12} through expanded sky coverage, refined spectroscopic reductions, and updated redshifts. The BOSS spectrographs provide moderate-resolution ($R\sim2,000$) spectra covering $\sim$3,600 to 10,400 \AA\ \cite{Smee2013}, well suited to broad ultraviolet absorption lines redshifted into the optical.

We used \texttt{baltools} \cite{Convolution_Neural_Network_PCA,DR16,Filbert2024,Martini2025} to identify \CIV\ outflow absorption troughs. The code considers quasars whose spectra cover the wavelength range out to 25,000 \kms\ blueward of the \CIV\ emission line, corresponding to $1.57<z_{\rm qso}<5$; most quasars in our sample have $z\sim2$ to 2.5. For these, \texttt{baltools} reconstructs each quasar continuum via $\chi^2$ minimization using principal component analysis (PCA) templates. We use the PCA templates of \cite{Brodzeller2023}, which we find improve performance relative to those used by \cite{DR16}, consistent with results reported for DESI data by \cite{Martini2025}. Broad ($>$450 \kms) absorption troughs are located blueward of the \CIV\ emission line and iteratively masked to improve the subsequent PCA fits; see \cite{Convolution_Neural_Network_PCA} and \cite{Martini2025} for full procedural details.

\begin{figure}
	\centering
	\includegraphics[width=0.7\textwidth]{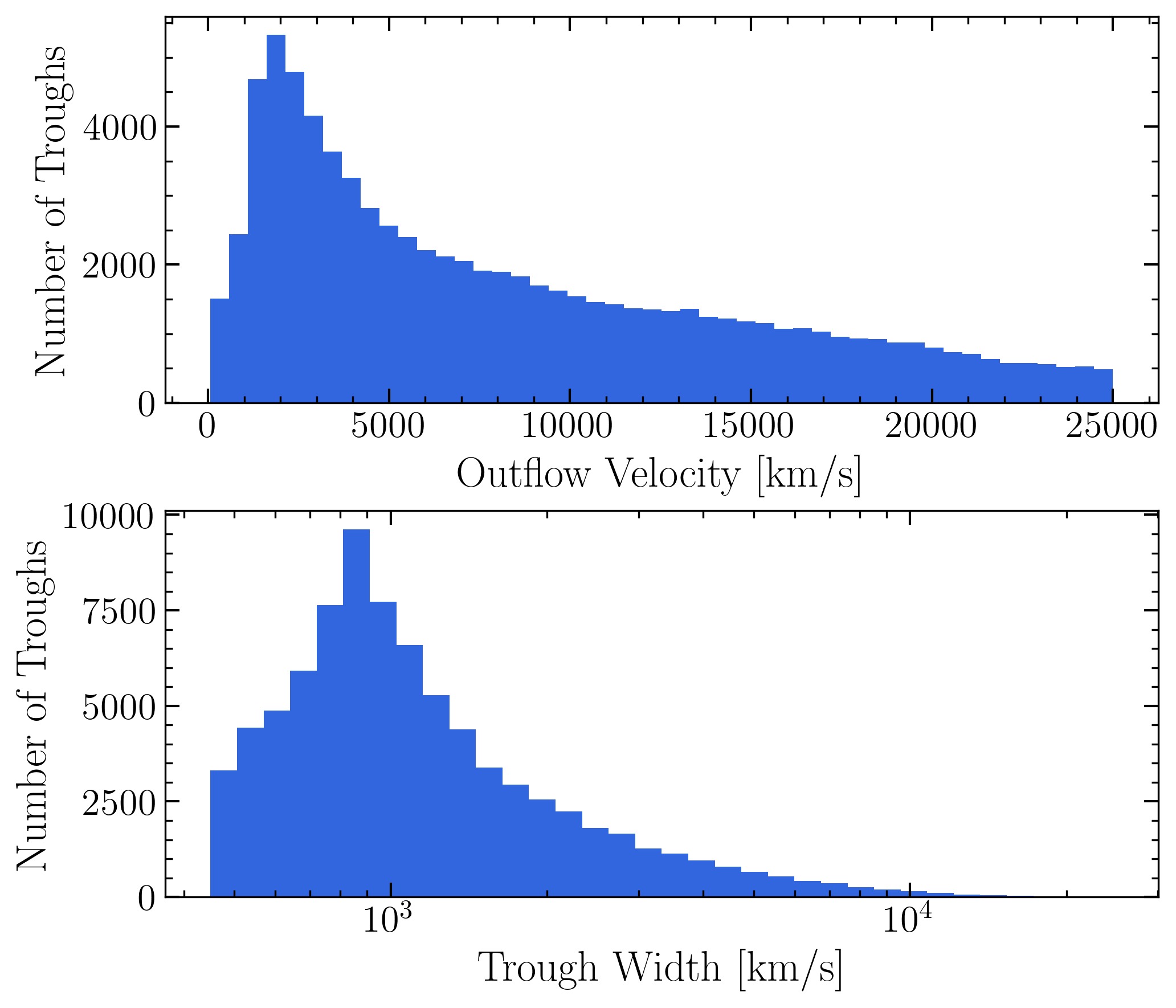}
	\caption{\textbf{Distributions of trough outflow velocity and width.}
	Outflow velocity, identified from the deepest point of the trough (top), and trough width (bottom), for the 81,388 troughs in the final sample.}
	\label{fig:s_trough_dist}
\end{figure}

Starting from 397,039 quasars satisfying the redshift requirement above, \texttt{baltools} identifies 159,690 \CIV\ troughs. We apply the quality and selection cuts listed in Table~\ref{tab:selection}, reducing the sample to a final 81,388 \CIV\ troughs, a 35 percent increase over the sample size used by \cite{Mas-Ribas2019}. Figure~\ref{fig:s_trough_dist} shows the resulting distributions of trough outflow velocity and width, which broadly match the DR12-based distributions of \cite{Mas-Ribas2019}.

\subsubsection*{Mock absorption trough construction}

To test whether the observed line locking signature could instead arise from stacking artifacts, we constructed a detailed mock spectral framework reproducing the statistical properties of real troughs in the absence of line locking.

Our mocks consist of two components: a \CIV\ doublet without line locking, $\mathcal{D}$, and a broad absorption trough, $\mathcal{T}$ (Fig.~\ref{fig:s_mock_example}). The doublet is modeled as two inverted Gaussians centered on $\lb$ and $\lr$,
\begin{equation}
\mathcal{D}(\lambda)=\exp\left[-b\,e^{\frac{-(\lambda-\lb)^2}{2\Delta^2}}-r\,e^{\frac{-(\lambda-\lr)^2}{2\Delta^2}}\right]~,
\label{eq:doublet}
\end{equation}
where $\Delta$ sets the width of the absorption features and $b$, $r$ are depth parameters for the blue and red components. The broad trough is modeled as an inverted boxcar centered at wavelength $c_t$,
\begin{equation}
\mathcal{T}(\lambda)=
\begin{cases}
\ell_b(\lambda), & \left[c_t-\tfrac{w}{2}\right]<\lambda\le\lb~,\\
\ell_r(\lambda), & \lb<\lambda<\left[c_t+\tfrac{w}{2}\right]~,\\
1, & \text{otherwise}~,
\end{cases}
\label{eq:trough}
\end{equation}
where $w$ is the trough width and $d$ is the trough depth. The term $\ell_r(\lambda)$ is
\begin{equation}
\ell_r(\lambda)=
\begin{cases}
1-d~, & \bar r\ge d~,\\
\dfrac{d}{c_t-\lb+\tfrac{w}{2}}\left[\lambda-\lb\right]+1-d~, & \bar r<d~,
\end{cases}
\label{eq:trough_mod}
\end{equation}
which connects the bottom of the trough at $\lb$ back to the continuum when the red line depth is less than the trough depth; $\ell_b(\lambda)$ is the equivalent expression for the blue side. We compare this choice against simpler alternatives in Fig.~\ref{fig:s_troughmodels} below.

Gaussian noise $\mathcal{N}$, with zero mean and standard deviation $n$, is added to the trough but not to the doublet, since $b$ and $r$ are measured from real (already noisy) spectra; $n$ represents both spectral noise and the physical variability of the trough bottom. $\mathcal{D}$ and $\mathcal{T}$ are evaluated over $\lambda\in[1500,1600]$ \AA, and the mock spectrum takes the lower value of $\mathcal{D}$ and $\mathcal{T}+\mathcal{N}$, except far from the doublet where only $\mathcal{T}+\mathcal{N}$ is used:
\begin{equation}
\mathcal{M}(\lambda)=
\begin{cases}
\mathcal{D}(\lambda)~, & \mathcal{D}(\lambda)<[\mathcal{T}(\lambda)+\mathcal{N}(\lambda)]\ \&\ \mathcal{D}(\lambda)<0.99~,\\
\mathcal{T}(\lambda)+\mathcal{N}(\lambda)~, & \text{otherwise}~,
\end{cases}
\label{eq:mockgen}
\end{equation}
with the restriction $\mathcal{M}=\mathcal{D}$ at $\lambda=\lb$ and $\lambda=\lr$, so that the mocks exactly match the doublet at the tips of the absorption lines.

\begin{figure}
	\centering
	\includegraphics[width=\textwidth]{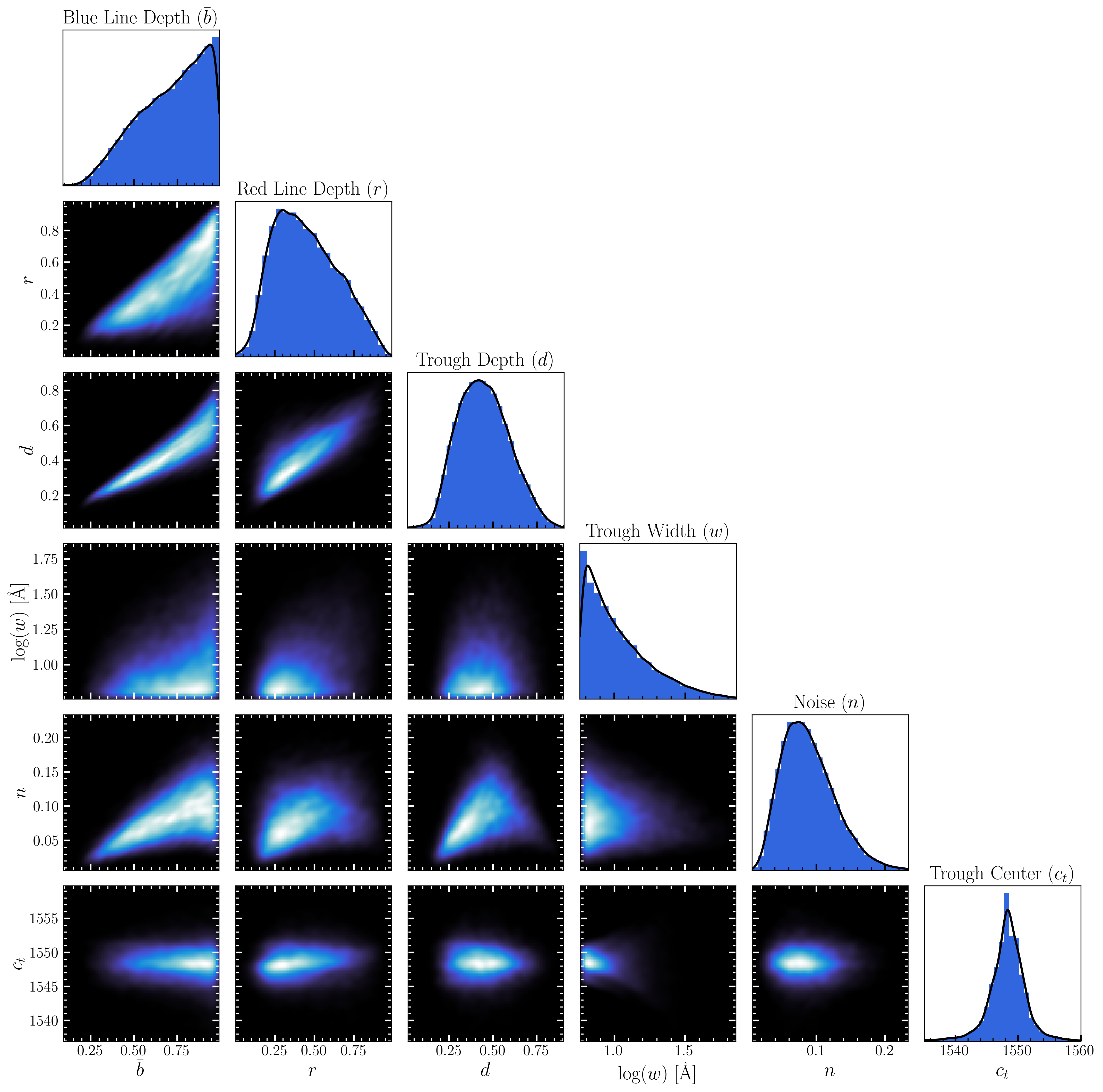}
	\caption{\textbf{Joint distribution of mock trough parameters.}
	Corner plot for the blue depth ($\bar b$), red depth ($\bar r$), trough depth ($d$), trough width ($w$), noise ($n$), and trough center ($c_t$) parameters describing the 29,332 SDSS troughs comprising the kernel density estimator (KDE) sub-sample. Blue histograms (diagonal panels) show measurements from the real SDSS data; black curves show the probability density functions inferred by the KDE. Off-diagonal panels show two-dimensional distributions between pairs of parameters, illustrating their correlations.}
	\label{fig:s_corner}
\end{figure}

\begin{figure}
	\centering
	\includegraphics[width=0.4\textwidth]{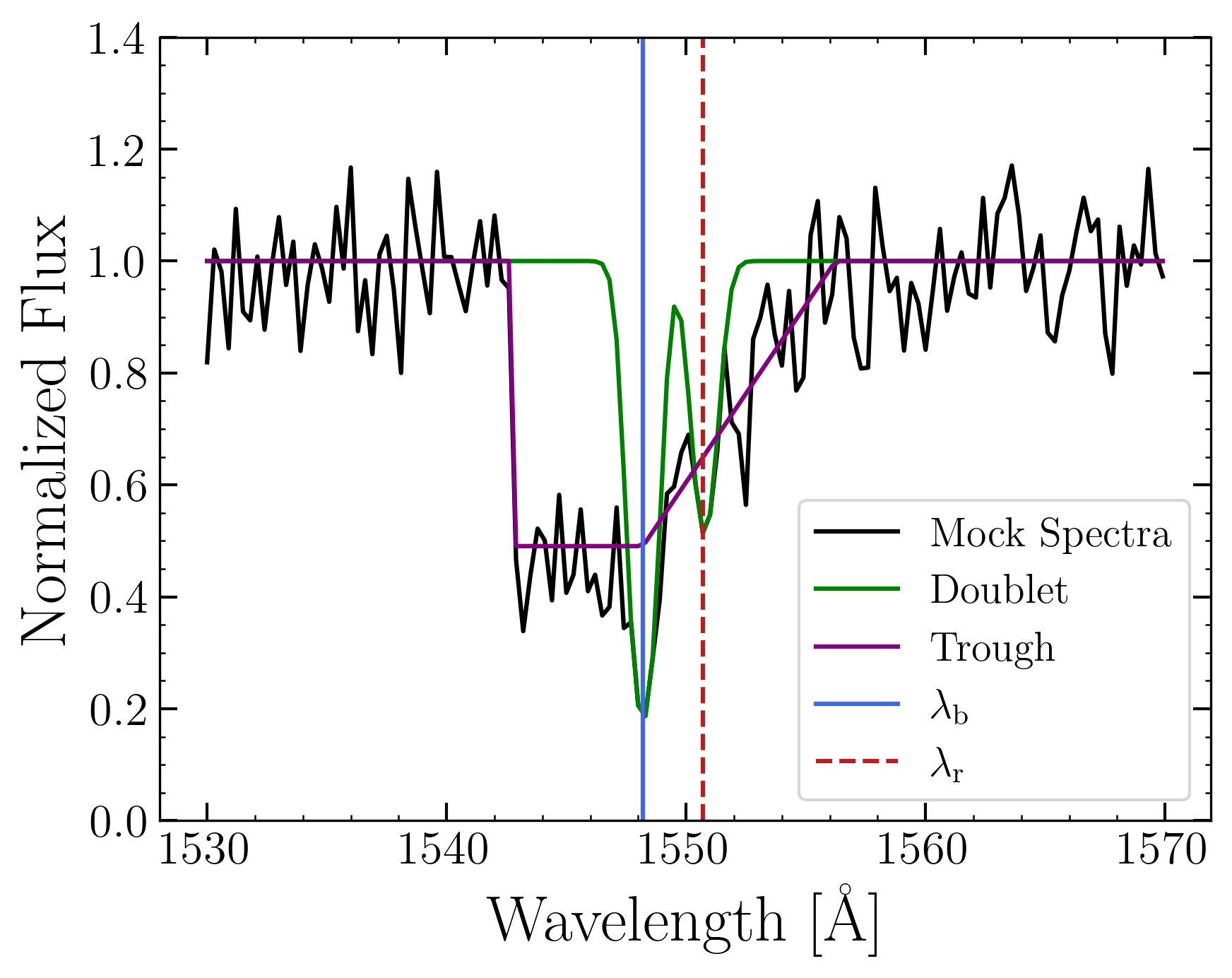}\hfill
	\includegraphics[width=0.56\textwidth]{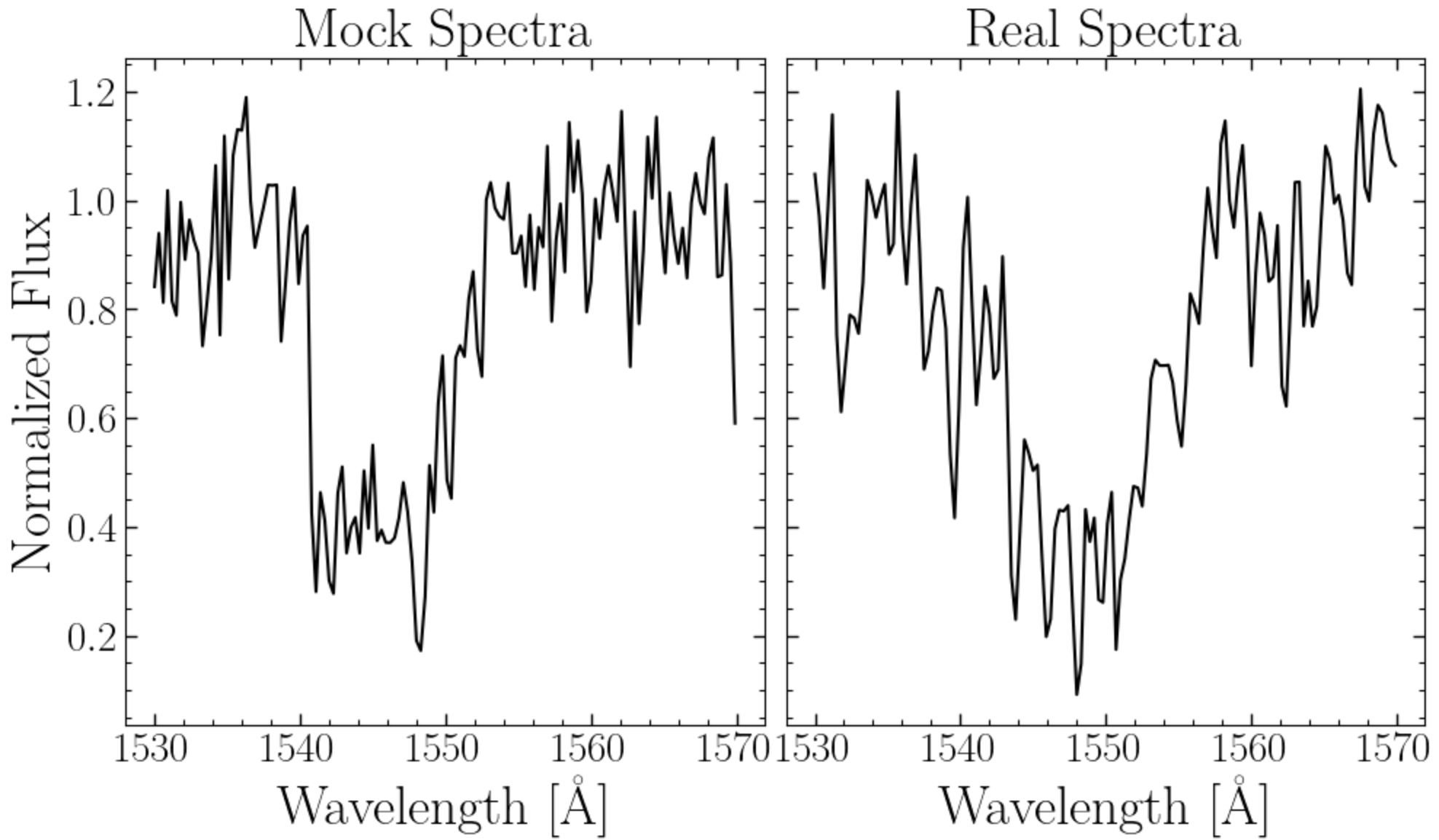}
	\caption{\textbf{Example mock absorption trough.}
	(Left) Example mock trough showing the doublet ($\mathcal{D}$, green) and trough ($\mathcal{T}$, purple) components, with vertical lines marking $\lb$ and $\lr$; this fiducial mock does not contain a line-locking feature. (Right) A mock absorption trough (left panel) compared to a real trough drawn from an SDSS/BOSS DR16 quasar spectrum (right panel), for illustration.}
	\label{fig:s_mock_example}
\end{figure}

Mock troughs are characterized by seven parameters: width $w$, depth $d$, center $c_t$, noise $n$, blue and red doublet depth parameters $b$ and $r$, and Gaussian width $\Delta$. We employ a six-parameter Gaussian KDE (bandwidth 1/8) to replicate the joint parameter space of the real data, including correlations between parameters (Fig.~\ref{fig:s_corner}). The bandwidth is chosen narrow to capture non-Gaussian behavior near the boundaries of the blue-line-depth and trough-width distributions. $\Delta=0.5$ \AA\ is fixed from Gaussian fits to stacked real spectra in \cite{Mas-Ribas2019,Mas-Ribas2019b}, since it cannot be extracted reliably from individual spectra and we find moderate variations around this value do not affect our results.

To construct the KDE, each spectrum is shifted to the trough rest frame using the redshift $z_{\rm trough}$ associated with the flux minimum, rebinned at 0.3 \AA\ (the BOSS spectral resolution, $\approx$70 \kms), and continuum-normalized to give $\mathcal{R}(\lambda)$. Trough boundaries $\LL$ and $\lu$ are determined from the velocities of the flux minimum and trough edges reported by \texttt{baltools}, giving $w=\lu-\LL$ and $c_t=(\lu+\LL)/2$. The normalized line depths $\bar b\equiv1-\mathcal{R}(\lb)$ and $\bar r\equiv1-\mathcal{R}(\lr)$ are converted to depth parameters via
\begin{equation}
b=\frac{e^\alpha\ln|1-\bar r|-\ln|1-\bar b|}{1-e^{2\alpha}}~,\qquad
r=\frac{e^\alpha\ln|1-\bar b|-\ln|1-\bar r|}{1-e^{2\alpha}}~,
\label{eq:depthparams}
\end{equation}
where $\alpha=-(\lb-\lr)^2/(2\Delta^2)$. Trough depth $d$ and noise $n$ are measured from a Gaussian-smoothed version of $\mathcal{R}(\lambda)$ over $[\LL,\lu]$, excluding the edge points and (for noise) the minimum flux point and its neighbors, following the procedure described in \cite{Mas-Ribas2019}. We require at least nine points after these exclusions, which imposes a minimum trough width of $\sim$1,200 \kms. Troughs below this width are excluded from the KDE only, leaving 29,322 troughs contributing to the KDE while all 81,388 troughs are retained for the outflow property analysis described below. We expect our results to hold for narrower troughs, since the width threshold is a measurement constraint rather than a physical one.

We generated 29,332 mock troughs by sampling the KDE. Figure~\ref{fig:s_mock_example} compares an example mock trough to a real DR16 trough. We find a small, systematic offset between mock and real trough-depth distributions, traced to the simplified slanted/boxcar trough shape; adding a constant offset of 0.0505 to the sampled trough depth (applied only to $d$, not to the doublet or triplet depth parameters) removes this offset and yields good agreement between the mock and real distributions across all parameters (Fig.~\ref{fig:s_paramcomp}; Table~\ref{tab:moments}).

\subsubsection*{Stacking methodology and validation tests}

We constructed composite (stacked) spectra following \cite{Mas-Ribas2019}. Each of the 29,332 mock spectra is centered at its flux minimum $\mathcal{M}(\lm)$ and assigned a weight
\begin{equation}
\omega_j=\frac{1}{n_j^2+0.1^2}~,
\label{eq:weight}
\end{equation}
where $n_j$ is the measured median noise of spectrum $j$ and the 0.1 term represents a 10 percent spectral variability floor that prevents very-low-noise spectra from disproportionately dominating the stack \cite{Mas-ribas2017}. The stacked spectrum is the weighted mean,
\begin{equation}
\mathcal{M}_{\rm stack}(\lambda_i)=\frac{1}{\Omega}\sum_{j=1}^{N}\omega_j\mathcal{M}_j(\lambda_i)~,
\label{eq:stack}
\end{equation}
with $\Omega=\sum_j\omega_j$ and $N=29{,}332$. The same procedure is applied to the real (PCA-normalized) spectra $\mathcal{R}$, using the inverse square of the median spectral signal-to-noise ratio in place of the noise estimate in Eqs.~\ref{eq:weight} and~\ref{eq:stack}.

Figure~1 (main text) shows that the doublet-only mock stack fails to reproduce the absorption seen at $\lv$ in the real data. From 50 independent bootstrap realizations of $\mathcal{M}_{\rm stack}$, the stacking algorithm aligns on the blue line ($\lm=\lb\pm0.3$ \AA) for 81.8$\pm$0.10 percent of troughs, on the red line ($\lm=\lr\pm0.3$ \AA) for 0.44$\pm$0.06 percent, and on noise for 17.7$\pm$0.21 percent. Widening the red-line contamination window to $\pm$0.6 or $\pm$0.9 \AA\ roughly doubles or triples this contamination fraction, but it remains below a few percent in all cases. We additionally tested three different trough-edge models (a strict boxcar, a fully slanted boxcar, and our fiducial mixed model; Fig.~\ref{fig:s_troughmodels}); all three give red-line contamination at the few-percent level (3 to 4 percent), and the fiducial model gives the best match to the real stack morphology (Fig.~\ref{fig:s_doubletstack}).

To construct line-locked mock troughs, we replace the doublet profile $\mathcal{D}(\lambda)$ with a triplet profile, $\tilde{\mathcal{D}}$, with components at $\lb$, $\lr$, and $\lv$ (see ``Mock triplet construction'' below). A mock sample built entirely from this triplet profile correctly produces absorption at $\lv$, but the resulting feature is more pronounced than in the real data. We find through visual inspection that a combination of roughly two thirds triplet and one third doublet mocks provides the best match (main text, Fig.~1), supporting a line locking fraction of approximately two thirds, consistent with the independent fraction measurement described in the main text.

\subsubsection*{Line locking fraction model}

We model the fraction of troughs exhibiting line locking using the two-cloud framework above. We assume a homogeneous absorbing clump of spatially uniform optical depth $\tau(\lambda)$ covering a fraction $f_c$ of a uniformly illuminating background source \cite{Arav1999}. The observed intensity for a single (non-line-locked) cloud is
\begin{equation}
I_1=I_0\left(1-f_c+f_c e^{-\tau(\lambda)}\right)~,\qquad
\tau(\lambda)=\tau_b e^{\frac{-(\lambda-\lb)^2}{2\Delta^2}}+\tau_r e^{\frac{-(\lambda-\lr)^2}{2\Delta^2}}~,
\label{eq:I1}
\end{equation}
with $\tau_b$, $\tau_r$ the line-center optical depths of the blue and red components. For a line-locked system, in which the second cloud's red component overlaps the first cloud's blue component,
\begin{equation}
I_{\ell\ell}=I_0\left(1-f_c+f_c e^{-\tau_{\ell\ell}(\lambda)}\right)~,\qquad
\tau_{\ell\ell}=(\tau_b+\tau_r)e^{\frac{-(\lambda-\lb)^2}{2\Delta^2}}+\tau_r e^{\frac{-(\lambda-\lr)^2}{2\Delta^2}}+\tau_b e^{\frac{-(\lambda-\lv)^2}{2\Delta^2}}~.
\label{eq:Ill}
\end{equation}
Contamination from misaligned (red-line) stacking is modeled as
\begin{equation}
I_c=I_0\left(1-f_c+f_c e^{-\tau_c(\lambda)}\right)~,\qquad
\tau_c(\lambda)=\tau_b e^{\frac{-(\lambda-\lv)^2}{2\Delta^2}}+\tau_r e^{\frac{-(\lambda-\lb)^2}{2\Delta^2}}~.
\label{eq:Ic}
\end{equation}

The total intensity in the real-data stack is
\begin{equation}
\mathcal{R}_{\rm stack}(\lambda)=\chi_c I_c+\chi_{\ell\ell}I_{\ell\ell}(\lambda)+(1-\chi_{\ell\ell}-\chi_c)I_1(\lambda)~,
\label{eq:stackmodel}
\end{equation}
where $\chi_{\ell\ell}$ and $\chi_c$ are the line-locked and contamination fractions. Writing $\beta(\lambda)=e^{-(\lambda-\lb)^2/2\Delta^2}$, $\alpha(\lambda)=e^{-(\lambda-\lr)^2/2\Delta^2}$, $\gamma(\lambda)=e^{-(\lambda-\lv)^2/2\Delta^2}$, $B=e^{-\tau_b}$, $R=e^{-\tau_r}$, and normalizing $I_0=1$, Eq.~\ref{eq:stackmodel} becomes
\begin{equation}
    \chi_{\ell\ell}f_c B^{\beta+\gamma}R^{\alpha+\beta}-(\chi_{\ell\ell}+\chi_c-1)f_c B^\beta R^\alpha
{}+\chi_c f_c B^\gamma R^\beta+1-f_c-\mathcal{R}_{\rm stack}=0~.
\end{equation}

There are six unknowns: $B$, $R$, $\chi_{\ell\ell}$, $\chi_c$, $f_c$, and $\Delta$. We evaluate this equation on the real stacked spectrum at $\lambda=\{\lb,\lr,\lv\}$, giving three equations solved numerically for $\chi_{\ell\ell}$, $B$, and $R$, while fixing $\chi_c$, $\Delta=0.5$ \AA, and $f_c$. We repeat this for $f_c=$ 80, 90, and 100 percent, and $\chi_c=$ 0 and 1 percent. The resulting line locking fraction is 60 to 72 percent (main text, Fig.~2), and the inferred optical depths sit at $\tau\sim1$ for both doublet components (Fig.~\ref{fig:s_opticaldepths}), as expected for line-driven winds at high blueshifted velocities \cite{Murray1995}.

\subsubsection*{Column density measurements}

Assuming an optically thin line with Doppler width $b_D=0.5$ \AA, the optical depth at line center is
\begin{equation}
\tau_0=\frac{\sqrt{\pi}e^2}{m_ec}\frac{N_{\rm CIV}f_{\rm osc}\lambda_0}{b_D}~,
\label{eq:tau0}
\end{equation}
where $N_{\rm CIV}$ is the \CIV\ column density and $f_{\rm osc}$ is the oscillator strength ($f_{\rm osc}=0.18999$ and 0.09520 for the blue and red \CIV\ components, respectively). Using the optical depths from the fraction-fitting procedure above, the implied average column density is $\sim1$ to $4\times10^{14}\,\cm^{-2}$ (Fig.~\ref{fig:s_density_cf}), essentially independent of whether $\tau_b$ or $\tau_r$ is used.

To examine the velocity dependence of the column density, troughs are sorted by outflow velocity (position of the flux minimum), and a window of 3,000 troughs is slid across the sorted sample, stacking the spectra within each window and extracting $\tau_r$ via the fitting procedure above. The resulting column density (Fig.~\ref{fig:s_density_vel}) increases with outflow velocity to a peak of $\sim4\times10^{14}\,\cm^{-2}$ at 3,500 \kms, then declines at higher velocities, consistent with increasing ionization parameter forcing carbon to a higher ionization state (\,C\,{\small V}\,) as the gas accelerates, and consistent with the equivalent-width trend reported by \cite{Mas-Ribas2019b}. A window size of 3,000 troughs balances smoothing against noise; smaller windows introduce features that are difficult to distinguish from noise, while larger windows reduce the sharpness of the observed peak.

Integrating the \CIV\ column density over the full velocity range gives a total $N_{\rm CIV}\approx6.3\times10^{16}\,\cm^{-2}$, which we can then use to find the corresponding hydrogen column:
\begin{equation}
N_{\rm H}=\frac{N_{\rm CIV}}{\chi_C\,\eta_{\rm CIV}}\approx4.8\times10^{20}\left(\frac{\chi_C}{3.26\times10^{-4}}\right)^{-1}\left(\frac{\eta_{\rm CIV}}{0.4}\right)^{-1}\cm^{-2}~,
\label{eq:NH}
\end{equation}
where $\chi_C$ is the carbon abundance, scaled to the solar value $\chi_{C,\odot}=3.26\times10^{-4}$ \cite{1989GeCoA..53..197A}, and $\eta_{\rm CIV}\equiv n_{\rm CIV}/n_C$ has a maximum value of 0.4 at ionization parameter $U=4\times10^{-3}$ \cite{1997ApJS..109..279H}.

If the gas-to-dust ratio in the outflow matched the Milky Way value, the continuum optical depth in the ultraviolet would be $\tau_d=\kappa_d m_p N_{\rm H}\approx0.8\,(\chi_C/\chi_{C,\odot})^{-1}(\eta_C/0.4)^{-1}$, where $\kappa_d\approx1000\,\cm^2\,\g^{-1}$ is the dust opacity near 1000 \AA. This is a factor of four larger than the observed optical depth, $\tau_d\approx0.1$ to 0.2 (Fig.~\ref{fig:s_dust}), suggesting an enhanced gas-to-dust ratio -- though we show below that this is an underestimate.

Our stacked spectra extend beyond the Lyman edge at 912 \AA\ (Fig.~\ref{fig:s_dust}), where the flux varies by less than 10 percent, giving $\tau_{\rm Ly\,edge}\lesssim0.1$. At the maximum value of $\eta_{\rm CIV}$, the predicted edge optical depth is $\tau_{\rm Ly\,edge}\approx4.4$, clearly excluded by the data. Since the observations fix $\tau_{\rm CIV}$ (and hence $N_{\rm CIV}$), reconciling this requires increasing the ionization parameter, which decreases $\eta_{\rm CIV}$ and correspondingly increases the inferred $N_{\rm H}$ via Eq.~\ref{eq:NH}. We find $U\approx0.2$ is required to satisfy the observed edge optical depth, giving
\begin{equation}
N_{\rm H}\approx5\times10^{21}\left(\frac{0.03}{\eta_{\rm CIV}}\right)^{-1}\cm^{-2}~,
\label{eq:NH2}
\end{equation}
with a corresponding $\tau_{\rm Ly\,edge}\approx0.1\,(\chi_C/4\chi_{C,\odot})^{-1}(\eta_C/0.03)^{-1}$, comfortably within observational limits assuming four times solar metallicity. For this larger $N_{\rm H}$, a Milky Way gas-to-dust ratio would predict $\tau_d\approx8$, clearly excluded by our spectra, which constrain $\tau_d$ to be at least 40 times smaller. We conclude that the outflowing gas has a gas-to-dust ratio at least 40 (and, accounting for the assumed metallicity, up to 160) times the Milky Way value -- that is, the outflow gas is essentially dust-free, as expected if the wind is launched from inside the dust sublimation radius.

This allows us to constrain the wind launch radius. The mass loss rate is $\dot M_w=4\pi r C_f m_p N_{\rm H}v_w\approx3.6\times10^{26}\,C_f\,(r/1.7\times10^{18}\,\cm)(v_w/20{,}000\,\kms)\,\g\,\sunit^{-1}$, scaled to the dust sublimation radius for $\lbol=10^{46}\,\ergs$ at sublimation temperature 1,500 K. Since the wind is radiatively driven, this mass loss rate must be smaller than $\lbol/(v_wc)$, requiring a launch radius $r_f\approx10^{17}\,\cm$, implying a terminal velocity of $\sim$30,000 \kms, consistent with observations and well inside the dust sublimation radius.

\subsubsection*{Mass outflow rate measurement}

We measured the flux absorbed by each \CIV\ trough by integrating the area between the PCA continuum and the observed spectrum (Fig.~\ref{fig:s_mdot_example}), applying the cuts in Table~\ref{tab:selection} except for the two flux-minimum conditions, since we wish to capture all absorption including possible noise contributions. This selection yields 116,428 troughs in 73,442 quasars.

To account for absorption from ionic species other than \CIV, we constructed a composite spectrum from 900 to 1,600 \AA\ rest-frame using SDSS-pipeline continua (rather than the PCA continua, which do not cover this full range), renormalized using a 50-pixel median boxcar smoothing after masking absorption features (Fig.~\ref{fig:s_extstack}, top panel). We then multiplied this renormalized absorption spectrum by the mean DR16 non-BAL quasar emission spectrum (selected to have null BAL probability), constructed using the same weighted-stacking procedure (Eqs.~\ref{eq:weight} and~\ref{eq:stack}) as above (Fig.~\ref{fig:s_extstack}, bottom panel). The difference between the resulting product and the non-BAL emission spectrum gives the average flux absorption at a given outflow velocity from all ions. Comparing this to the known \CIV\ absorption then gives the average \CIV\ fractional contribution to total absorption as a function of velocity (Fig.~\ref{fig:s_civratio}), found to range between approximately 11 and 17 percent.

The total absorbed luminosity along each quasar sightline is $L_{\rm trough,los}=4\pi D_L^2 F_{\rm trough,los}$, where $F_{\rm trough,los}$ is obtained by applying the velocity-dependent \CIV-to-total absorption ratio above. We estimate the covering fraction of the central engine by the outflow, $C_f$, from the fraction of quality-selected DR16 quasars showing at least one outflow trough: of 295,690 quasars passing our signal-quality cuts, 112,789 (38 percent) show outflow absorption features, giving $C_f=0.38$ (this is twice the covering fraction obtained when considering only strictly broad, $>$2,000 \kms, troughs; \cite{Hewett2003,Gibson2009}). The total absorbed luminosity for each trough is then $L_{\rm trough}=C_f L_{\rm trough,los}$, and the mass loss rate follows from momentum conservation under the assumption of single scattering,
\begin{equation}
\frac{L_{\rm trough}}{c}=\dot M v_{\rm trough}~.
\label{eq:momcons}
\end{equation}
Each BAL quasar has on average 1.6 troughs. The resulting distributions of mass outflow rate, quasar bolometric luminosity, absorbed luminosity fraction, mass outflow to accretion rate ratio, and kinetic to bolometric luminosity ratio are shown in main text Fig.~3. The mass accretion rate is $\dot M_{\rm acc}=\lbol/(\eta c^2)$ with mass-to-energy efficiency $\eta=0.1$.

\subsubsection*{Two-population clustering and dust reddening}

The distributions in main text Fig.~3 show clear bimodality (except for bolometric luminosity and kinetic-to-bolometric ratio). We applied the DBSCAN density-based clustering algorithm \cite{Ester1996} to the three-dimensional space of outflow mass rate, absorbed luminosity fraction, and mass ratio. We tested DBSCAN against K-means, spectral clustering, and HDBSCAN \cite{hdbscan}, both on the raw data and after dimensionality reduction via PCA and t-SNE \cite{tsne}; all methods consistently identify the same two cluster cores, but DBSCAN requires the least hyperparameter tuning and, importantly, labels ambiguous points as unassigned rather than forcing them into one cluster or the other.

The low-mass-outflow-rate population is composed mostly of one or two narrow ($<$2,000 \kms) troughs in quasars with BAL probability near 0.9, with 90 percent of these troughs at outflow velocities above 8,500 \kms\ (Fig.~\ref{fig:s_populations}). The high-mass-outflow-rate population extends to larger trough widths and lower velocities. Stacking the real data for each population separately (Fig.~\ref{fig:s_dust}) shows that line locking is clearly visible in the high-mass-outflow-rate stack but not in the low-mass-outflow-rate stack.

To assess dust content, we computed the mean quasar emission spectrum for each population, masking the \CIV, \NV, \HI, \OVI, and \SiIV\ trough regions to mitigate their effect on the continuum (Fig.~\ref{fig:s_dust}, top panel). The high-mass-outflow-rate stack is notably redder than the non-BAL quasar continuum, indicating substantial dust absorption \cite{Weymann1991,Reichard2003}, while the low-mass-outflow-rate stack is very similar to the non-BAL continuum. Quantifying this via the extinction $A_\lambda=-2.5\log_{10}(F_{\rm cluster}/F_{\rm non\text{-}BAL})$ (Fig.~\ref{fig:s_dust}, middle panel), and integrating over the continuum, the low-mass population absorbs 4.61 percent of the non-BAL flux and the high-mass population absorbs 9.63 percent. Applying our cover fraction $C_f=0.38$, this corresponds to dust absorption of 1.75 and 3.66 percent of $\lbol$ for the two populations, respectively, a factor of 5.83 and 2.44 times larger than the corresponding line absorption fractions. The bottom panel of Fig.~\ref{fig:s_dust} shows the fraction of spectra contributing to each stack as a function of wavelength, illustrating the effect of trough masking.

\subsubsection*{CAK wind dynamics and line locking}

We model the wind acceleration using a simplified, steady-state version of the standard Castor, Abbott \& Klein (CAK) formalism \cite{CAK1975}, assuming a radial flow emerging from a rotationally supported disk and driven primarily by line absorption. The momentum equation is \cite{Murray1995}
\begin{equation}
v\frac{dv}{dr}=-\frac{GM_\bullet}{r^2}+\frac{v_\phi^2}{r}+\frac{1}{\rho}\frac{dP}{dr}+g_\ell+g_e~,
\label{eq:momeq1}
\end{equation}
where $M_\bullet$ is the black hole mass, $v_\phi$ the angular velocity, $P$ and $\rho$ the gas pressure and density, and $g_\ell$, $g_e$ the line-driving and electron-scattering acceleration terms. Neglecting the gas pressure term (valid in the supersonic limit) and assuming single-scattering for both driving terms, $g_e=\sigma_T\lbol/(4\pi m_p c r^2)$, and $g_\ell=g_e\mathcal{M}_{\rm CAK}(t)$, where the force multiplier is
\begin{equation}
\mathcal{M}_{\rm CAK}(t)\equiv k\left(\frac{m_p}{\rho v_{\rm th}\sigma_T}\left|\frac{dv}{dr}\right|\right)^\alpha\left(\frac{10^{-11}n_e}{W}\right)^\xi~,
\label{eq:cak}
\end{equation}
with $v_{\rm th}$ the wind thermal velocity, $n_e$ the electron density, $W$ the geometric dilution factor, and $k$, $\delta$, $\alpha$ constants. Using mass conservation, $\dot M=4\pi r^2\rho v$, to eliminate $\rho$, and the Eddington ratio $\Gamma_{\rm Edd}\equiv\sigma_T\lbol/(4\pi m_p cGM_\bullet)$, and using $rv_\phi^2=\ell^2/r$ with $\ell\equiv rv_\phi=$ const., together with the boundary condition $v\,dv/dr|_{r=r_0}\approx0$ at the launch radius $r_0$ (which gives $\ell^2=GM_\bullet r_0(1-\Gamma_{\rm Edd})$), the momentum equation reduces to
\begin{equation}
r^2v\frac{dv}{dr}=-GM_\bullet\left[(1-\Gamma_{\rm Edd})\left(1-\frac{r_0}{r}\right)-\Gamma_{\rm Edd}k\left(\frac{10^{-11}n_e}{W}\right)^\xi\left(\frac{4\pi m_p r^2}{v_{\rm th}\sigma_T\dot M}v\left|\frac{dv}{dr}\right|\right)^\alpha\right]~.
\label{eq:momeqfinal}
\end{equation}
This nonlinear first-order equation has a unique outflow solution passing through a critical point, approximated by \cite{Murray1995} as
\begin{equation}
v(r)=v_\infty\left(1-\frac{r_0}{r}\right)^\beta~,\qquad
\beta=\frac{1+\alpha}{2\alpha-\xi}~,
\label{eq:betalaw}
\end{equation}
\begin{equation}
v_\infty\approx\sqrt{\frac{\alpha}{2\beta(1-\alpha)}}\,v_{\rm esc}=\sqrt{\frac{\alpha}{\beta(1-\alpha)}\frac{GM_\bullet(1-\Gamma_{\rm Edd})}{r_0}}~,
\label{eq:vinf}
\end{equation}
indicating that line driving can accelerate winds to a couple of times their initial escape velocity. For $M_\bullet=10^8\,M_\odot$, $\lbol=10^{46}\,\ergs$, and fiducial $\alpha=0.8$ \cite{Murray1995} and $\xi=10^{-3}$, matching the observed terminal velocities of order 30,000 \kms\ requires $r_0\approx10^{15}(M_\bullet/10^8M_\odot)(v_\infty/30{,}000\,\kms)^{-2}$ cm and $\beta\sim1.1$ (main text Fig.~4).

We use this acceleration profile, together with the line locking condition (Eq.~\ref{eq:llcondition}), to determine the velocity range over which line locking can be sustained, parameterized as
\begin{equation}
\epsilon(v)=\frac{a_2(v+\delta)-a_1(v-\delta)}{|a_2(v+\delta)|}~,
\label{eq:epsv}
\end{equation}
with $\delta\equiv497/2$ \kms\ half the \CIV\ doublet velocity separation. To estimate $\epsilon$ in the real data, we consider three physical factors limiting the fractional reduction in $|a_2|$ during shadowing: (i) the importance of line driving relative to other forces, $g_\ell/g_{\rm other}$; (ii) the contribution of \CIV\ to the total line-driving force, $F_{\rm CIV}/F_{\rm other}$; and (iii) the fraction of \CIV\ flux actually masked during a lock, $F_{\rm masked}$, giving
\begin{equation}
\epsilon=\frac{g_\ell}{g_{\rm other}}\times\frac{F_{\rm CIV}}{F_{\rm other}}\times F_{\rm masked}~.
\label{eq:epsilon}
\end{equation}
We take $g_\ell/g_{\rm other}\sim1$ (line driving as the only force acting), $F_{\rm CIV}/F_{\rm other}\approx0.15$ (Fig.~\ref{fig:s_civratio}), and estimate $F_{\rm masked}$ by constructing single-cloud and line-locked two-cloud \CIV\ absorption profiles at covering fraction $f_c=0.9$ using the fitted optical depths (Fig.~\ref{fig:s_masking}): the integrated absorption areas are $A_1=1.44$ and $A_{\ell\ell}=2.56$, giving $F_{\rm masked}=(A_{\ell\ell}-A_1)/A_1\approx0.2$. Combining these gives $\epsilon\approx0.03$, which intercepts the acceleration curve at $\sim$7,000 \kms\ and enters the disallowed (red) region at $\sim$10,000 \kms\ (main text Fig.~4).

Composite spectra in narrow velocity bins (main text Fig.~5) show line locking signatures well above and below 7,000 \kms\ and above 10,000 \kms, a wider range than this simple estimate of active locking alone predicts. We account for this using the Sobolev length $\ell_{\rm Sob}=v_{\rm th}(dv/dr)^{-1}$ and corresponding cloud dispersion timescale $t_{\rm dis}\approx(dv/dr)^{-1}$, which, using Eq.~\ref{eq:betalaw}, can be written
\begin{equation}
t_{\rm dis}(x)=\frac{r_0}{\beta v_\infty}\frac{x^{(1-\beta)/\beta}}{\left(1-x^{1/\beta}\right)^2}~,\qquad x=v/v_\infty~.
\label{eq:tdis}
\end{equation}
At $v=10{,}000$ \kms\ this gives $t_{\rm dis}=8.4\times10^5$ s. The corresponding travel time between two velocities is
\begin{equation}
t_{\rm travel}=\frac{r_0}{\beta v_\infty}\int_{x_1}^{x_2}\frac{dx}{x^2\left(x^{-1/(2\beta)}-x^{1/(2\beta)}\right)^2}~.
\label{eq:ttravel}
\end{equation}
For $\beta=1.1$, $v_\infty=30{,}000$ \kms, and $r_0=10^{15}$ cm, gas leaving the actively line-locked region at $\sim$10,000 \kms\ can travel to $\sim$18,500 \kms\ before $t_{\rm travel}\ge t_{\rm dis}$. Thus, while active line locking occurs only over a narrow velocity range, the locked velocity configuration can be passively maintained over a substantially larger range before dispersing, consistent with the observed persistence of line locking at our largest velocity bin in the main text Fig.~5.

\subsubsection*{Mock triplet construction}

\begin{figure}
	\centering
	\includegraphics[width=\textwidth]{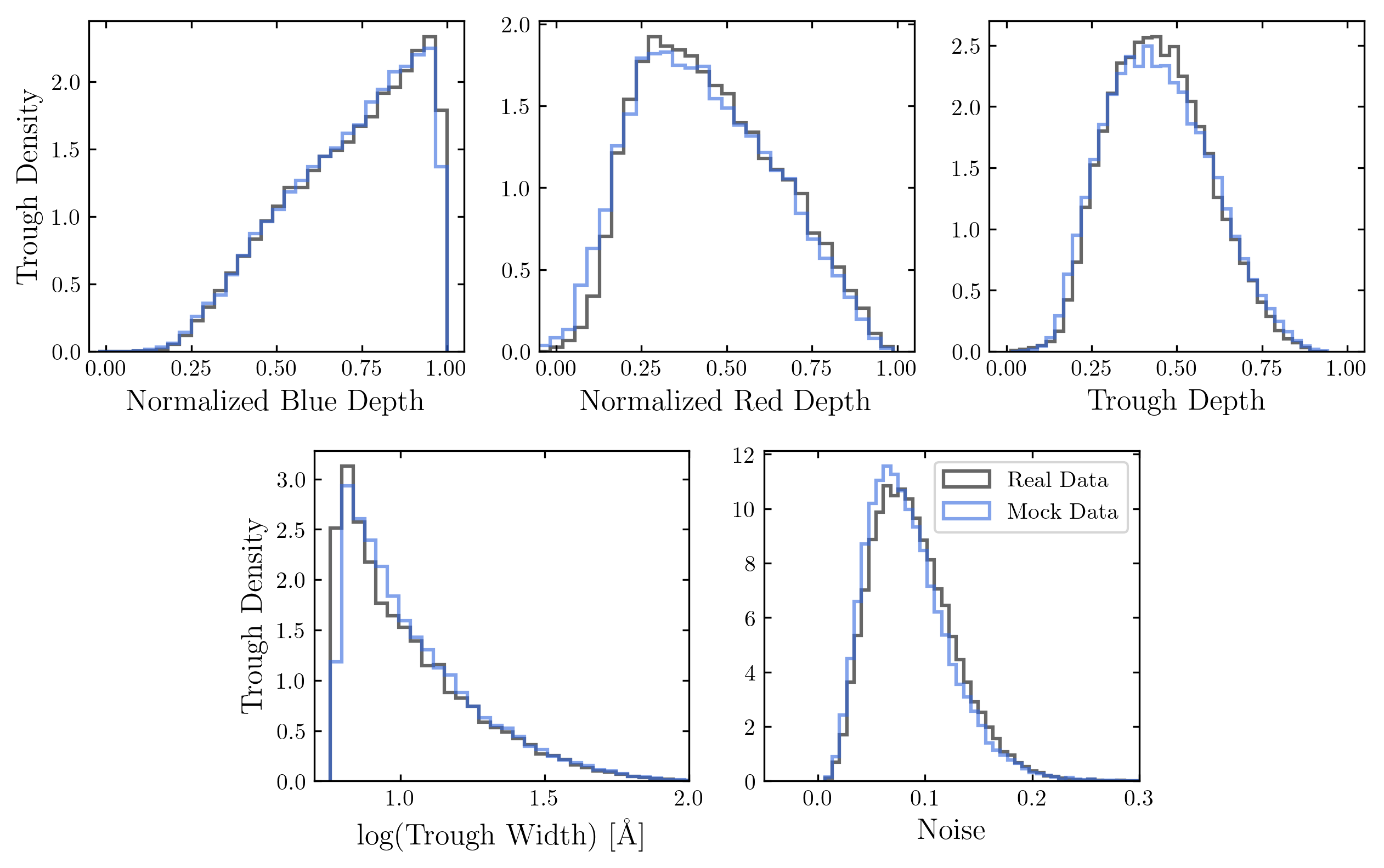}
	\caption{\textbf{Comparison of real and mock trough parameter distributions (doublet-only mocks).}
	Comparison of normalized blue depth, normalized red depth, trough depth, log trough width, and noise between the real (black) and mock (blue) samples, after applying the depth offset described in Materials and Methods. Broad agreement between the distributions indicates that the mocks successfully replicate the properties of the real data; summary statistics are given in Table~\ref{tab:moments}.}
	\label{fig:s_paramcomp}
\end{figure}

The line-locked mock model is
\begin{equation}
\tilde{\mathcal{M}}(\lambda)=
\begin{cases}
\tilde{\mathcal{D}}(\lambda)~, & \tilde{\mathcal{D}}(\lambda)<[\tilde{\mathcal{T}}(\lambda)+\mathcal{N}(\lambda)]\ \&\ \tilde{\mathcal{D}}(\lambda)<0.99~,\\
\tilde{\mathcal{T}}(\lambda)+\mathcal{N}(\lambda)~, & \text{otherwise}~,
\end{cases}
\label{eq:tripletmodel}
\end{equation}
where the triplet absorption profile is
\begin{equation}
\tilde{\mathcal{D}}=\exp\left[-b\,e^{\frac{-(\lambda-\lb)^2}{2\Delta^2}}-r\,e^{\frac{-(\lambda-\lr)^2}{2\Delta^2}}-v\,e^{\frac{-(\lambda-\lv)^2}{2\Delta^2}}\right]~,
\label{eq:triplet}
\end{equation}
with $b$, $r$, $v$ the depth parameters of the blue, red, and line-locked components. Constructing a triplet mock trough requires an eighth parameter, the locked-line depth $\bar v=1-\mathcal{R}(\lv)$, measured from the real data and added to the KDE described above. Converting $\bar b$, $\bar r$, $\bar v$ to depth parameters $b$, $r$, $v$ requires inverting the $3\times3$ system obtained by evaluating Eq.~\ref{eq:triplet} at $\lb$, $\lr$, $\lv$; the inverted relation is
\begin{equation}
\begin{bmatrix}b\\r\\v\end{bmatrix}=
\frac{1}{e^{2\beta}+2e^{2\alpha}(1-e^\beta)-1}
\begin{bmatrix}
1-e^{2\beta} & e^\alpha(e^\beta-1) & e^\alpha(e^\beta-1)\\
e^\alpha(e^\beta-1) & 1-e^{2\alpha} & e^{2\alpha}-e^\beta\\
e^\alpha(e^\beta-1) & e^{2\alpha}-e^\beta & 1-e^{2\alpha}
\end{bmatrix}
\begin{bmatrix}\ln|1-\bar b|\\ \ln|1-\bar r|\\ \ln|1-\bar v|\end{bmatrix}~,
\label{eq:tripletmatrix}
\end{equation}
where $\alpha=-(\lr-\lb)^2/(2\Delta^2)$ and $\beta=-(\lr-\lv)^2/(2\Delta^2)$. We built 29,332 triplet mock troughs (applying the same depth offset described above) using a mixture of two thirds triplet and one third doublet troughs, matching the proportion found to best reproduce the real stack (main text Fig.~1). Figure~\ref{fig:s_paramcomp_triplet} compares the resulting mock and real parameter distributions, showing good overall agreement, similar to the doublet-only case (Fig.~\ref{fig:s_paramcomp}).


\subsection*{Supplementary Text}

\subsubsection*{Extended background on AGN feedback}

Beyond the galaxy-scale simulations discussed in the main text, there is substantial direct observational evidence that quasars affect their host galaxies \cite{1993MNRAS.263..323T,2011ApJ...733L..16S,2012ARA&A..50..455F,2013ARA&A..51..511K,Chen2022}, the surrounding circumgalactic medium \cite{2013MNRAS.436.2576L,2025ApJ...989..230V}, and even, indirectly, the intergalactic medium \cite{2010MNRAS.405.1025M}. For example, \cite{2011ApJ...733L..16S} report p-Cygni profiles from OH molecules in nearby ultra-luminous infrared galaxies, evidence of outflows exceeding 1,000 \kms\ with mass loss rates of up to 1,200 $M_\odot\,\mathrm{yr}^{-1}$, comparable to or exceeding the star formation rate, with more powerful AGN hosting more massive, higher-velocity outflows. Separately, \cite{2013MNRAS.430.2327L} present integral field spectroscopy of [O~{\small III}] emission extending to 15 kpc with a median outflow velocity of 760 \kms, used to rule out starburst-driven winds in favor of an AGN origin. Indirect evidence comes from the scaling relations between black hole mass and host bulge stellar mass \cite{1998AJ....115.2285M} and stellar velocity dispersion \cite{2000ApJ...539L..13G}, which motivated early black hole feedback theories \cite{1998A&A...331L...1S} and later momentum-driven models \cite{2002MNRAS.329L..18F,2003ApJ...596L..27K,2005ApJ...618..569M}. A further hint comes from the relation between stellar and dark matter halo mass \cite{2002ApJ...569..101M,2010ApJ...717..379B,2010ApJ...710..903M}, likely shaped by the powerful jets that operate in galaxy groups and clusters, where feedback from large-scale radio jets is unambiguously responsible for offsetting the cooling of hot intracluster gas \cite{2007ARA&A..45..117M}.

Early galaxy formation models lacking any feedback mechanism predicted far too many low-mass, actively star-forming galaxies, the ``overcooling problem'' \cite{1978MNRAS.183..341W,1991ApJ...367...45C,1991ApJ...379...52W}, while simultaneously predicting too many massive galaxies. Semi-analytic models that track supermassive black hole growth \cite{2000MNRAS.311..576K}, and later models motivated by the interaction of radio jets with intracluster gas \cite{2000ApJ...534L.135M,2006MNRAS.365...11C,2006MNRAS.370..645B}, reproduce the observed cutoff in the space density of massive galaxies but do not model feedback from AGN in their most luminous, quasar phase. Numerical simulations lacking stellar and AGN feedback similarly overcool \cite{1997ApJ...478...13N}; resolving stellar feedback removes this overcooling in small cosmological volumes \cite{2011MNRAS.417..950H,2012MNRAS.421.3522H,2014MNRAS.445..581H,2014MNRAS.437.3529K,2015ApJ...804...18A}, but such simulations still overproduce stars in more massive halos unless AGN feedback is also included, motivating its incorporation in the cosmological simulations discussed in the main text.

Proposed physical mechanisms for AGN feedback span a wide range of scales: heating of host-galaxy gas by AGN radiation \cite{2005Natur.433..604D} and radiation pressure on dust grains in the host interstellar medium \cite{2005ApJ...618..569M,Ishibashi2024} operate on kiloparsec scales, while sub-parsec-scale winds have been proposed to be driven by unspecified means \cite{1998A&A...331L...1S}, by radiation pressure on resonance lines \cite{Murray1995,2000ApJ...543..686P,Nomura2016}, by electron scattering in Eddington or super-Eddington sources \cite{2003ApJ...596L..27K}, thermally \cite{Netzer2007}, or by magnetic fields \cite{1982MNRAS.199..883B,Emmering1992}; neutron-driven winds have also been proposed on somewhat larger scales \cite{1989ApJ...341L..33S}. This work addresses which of these mechanisms dominates in observed broad absorption line winds.

\subsubsection*{Interpretation of the line locking fraction}

A line locking fraction below 100 percent does not necessarily indicate that radiative acceleration is subdominant in the remaining outflows. Ionization stratification within an outflow could deplete \CIV\ in some regions while other ionic species continue to be radiatively accelerated, breaking the \CIV\ line locking signature without the underlying driving mechanism becoming subdominant. Signal-to-noise limitations in individual spectra may also cause us to underestimate the true line locking fraction, since weaker line-locked systems may not produce detectable absorption at the expected position in the stacks. Our measured fraction of 60 to 70 percent should therefore be regarded as a lower limit on the prevalence of line-driven outflows.

Combining this fraction with the 20 to 40 percent of luminous quasars that exhibit broad absorption troughs \cite{2023ApJ...952...44B} implies that radiative line driving observably operates in approximately 25 percent of all quasars, again a lower limit, since some systems may be driven by radiation pressure without producing the specific velocity alignment required to detect line locking.

\subsubsection*{Sensitivity to the mock trough model}

\begin{figure}
	\centering
	\includegraphics[width=0.95\textwidth]{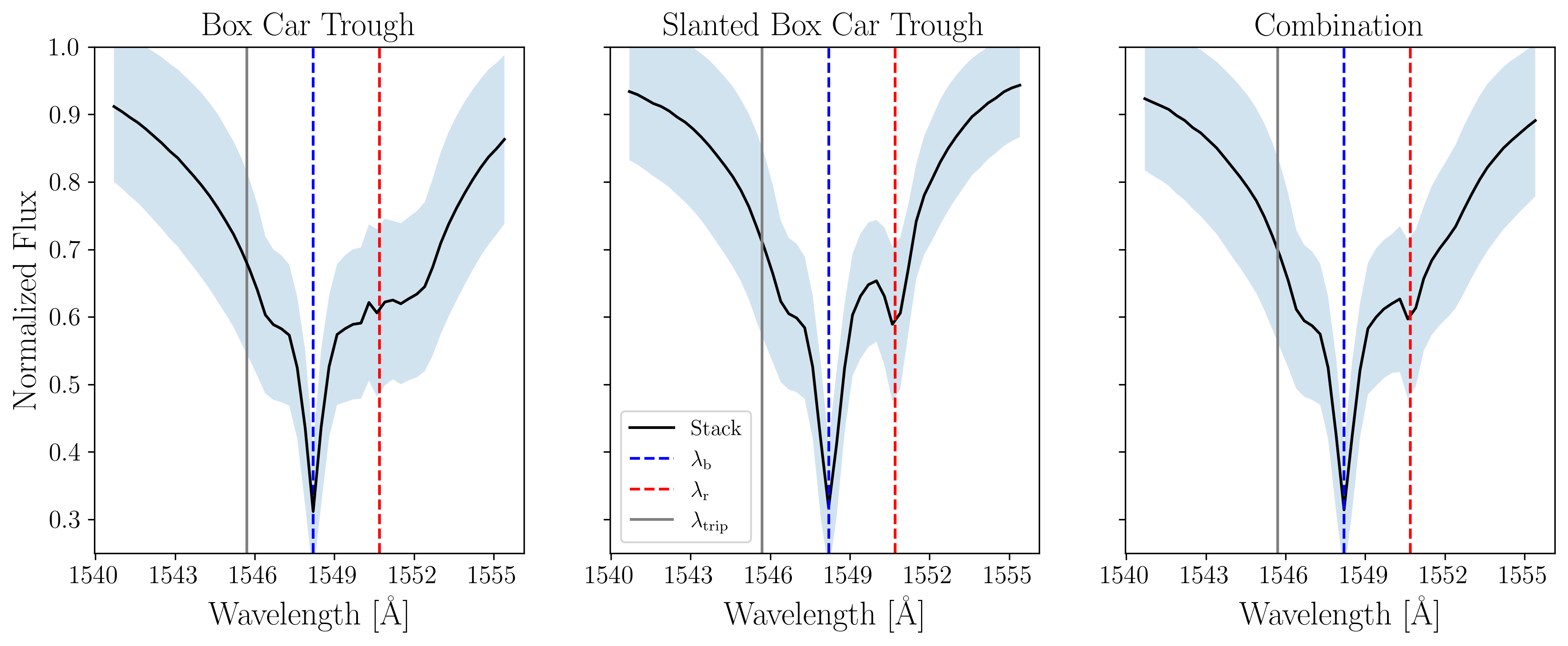}
	\caption{\textbf{Effect of trough-edge model choice on the stacked spectrum.}
	Stacks resulting from three trough models: an inverted boxcar with no slanted modification (left), a boxcar with the slanted modification always applied on the red side (middle), and a combination of the two depending on whether the red line depth is less than or greater than the trough depth, our fiducial model (right).}
	\label{fig:s_troughmodels}
\end{figure}

\begin{figure}
	\centering
	\includegraphics[width=0.55\textwidth]{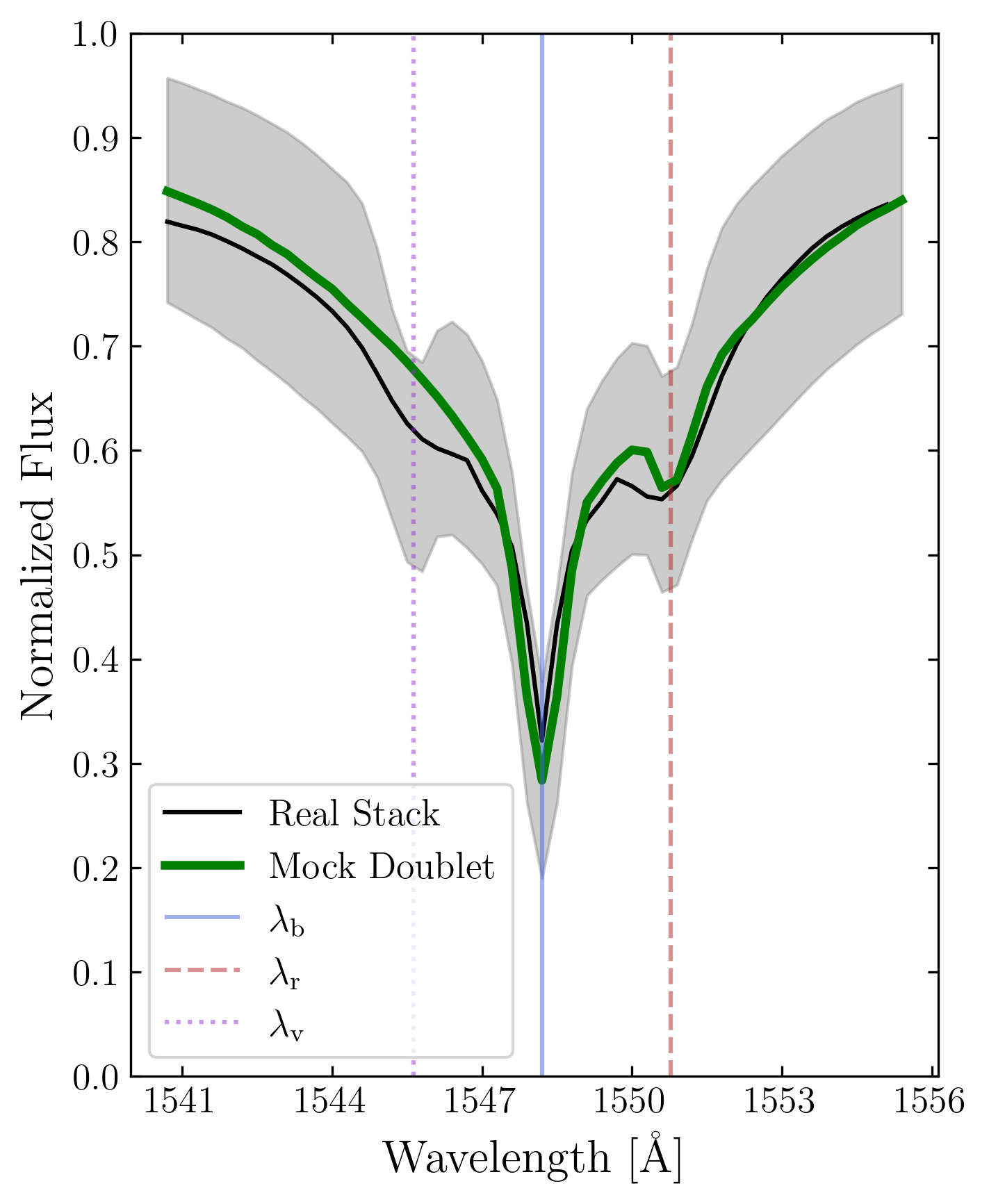}
	\caption{\textbf{Doublet-only mock stack compared to the real data stack.}
	Same comparison as main text Fig.~1, showing only the doublet mock (green) and real data (black) stacks, without the triplet or combined mocks. The shaded region shows half the weighted standard deviation of the real stack.}
	\label{fig:s_doubletstack}
\end{figure}

\begin{figure}
	\centering
	\includegraphics[width=0.6\textwidth]{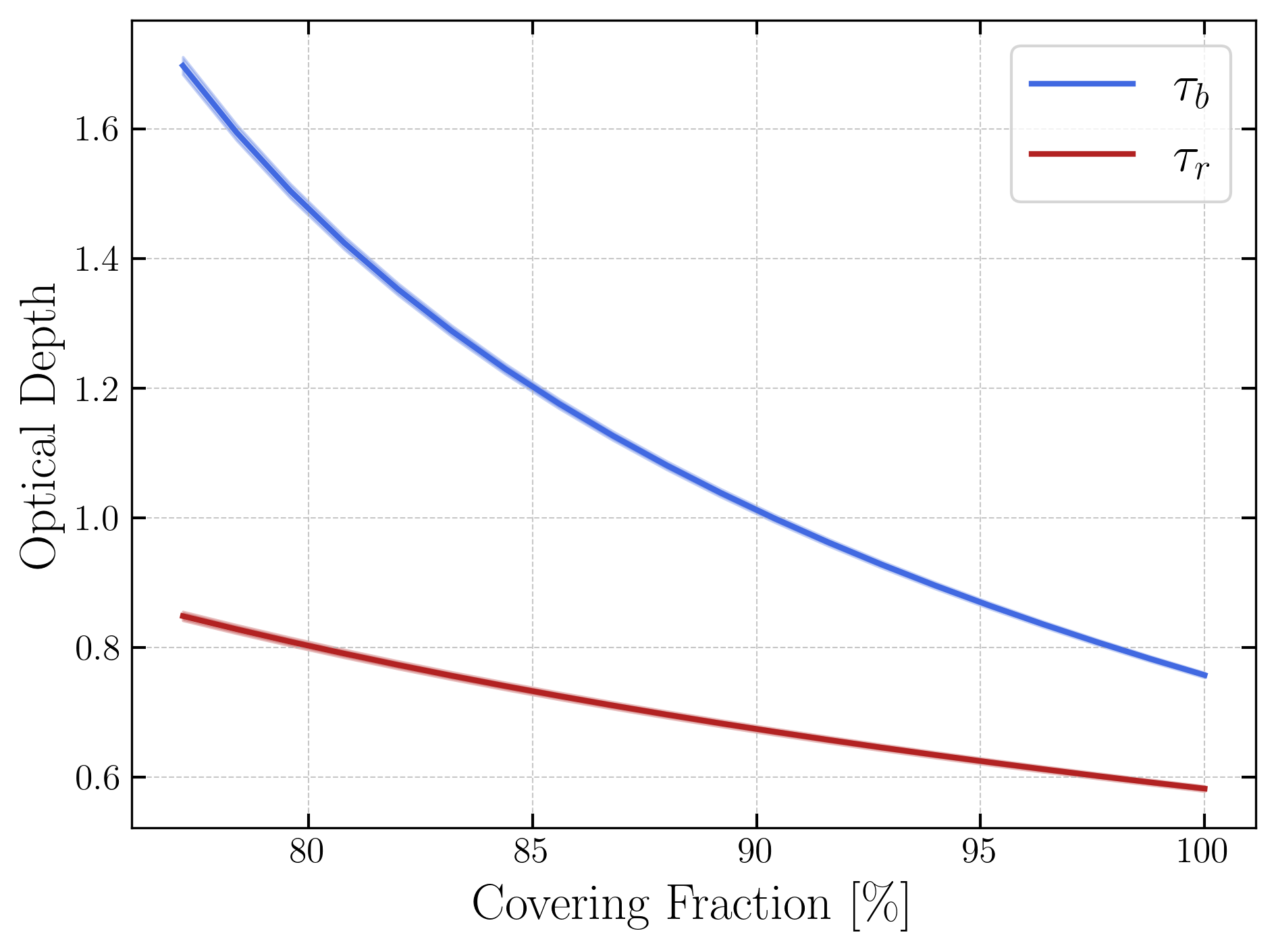}
	\caption{\textbf{Fitted \CIV\ optical depths as a function of covering fraction.}
	Optical depths of the blue and red \CIV\ absorption components, derived from the line-locking-fraction model (Eq.~\ref{eq:stackmodel}) evaluated on the full real trough sample.}
	\label{fig:s_opticaldepths}
\end{figure}

\begin{figure}
	\centering
	\includegraphics[width=0.6\textwidth]{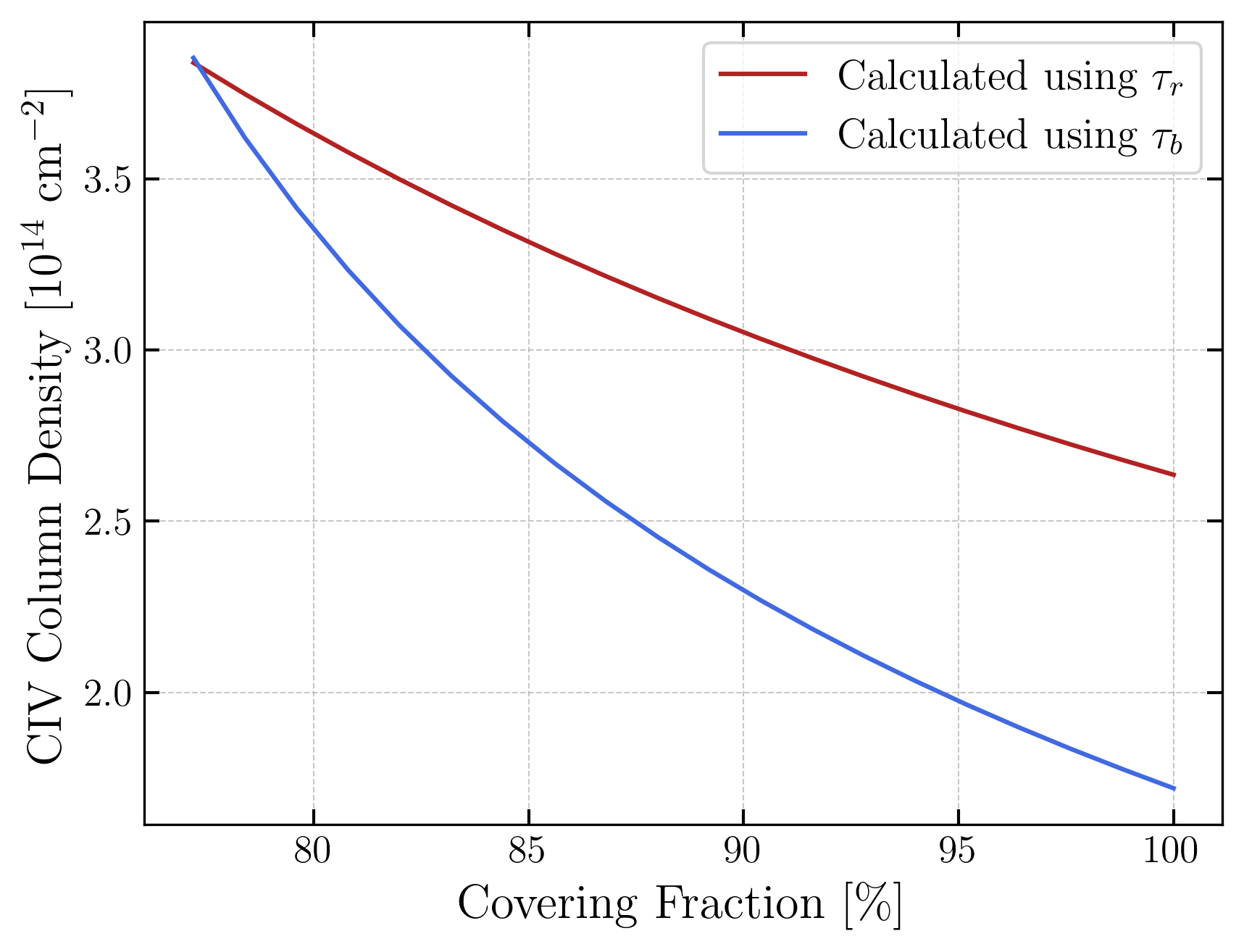}
	\caption{\textbf{\CIV\ column density as a function of covering fraction.}
	Lower limit on the \CIV\ column density derived from the optical depths in Fig.~\ref{fig:s_opticaldepths} via Eq.~\ref{eq:tau0}. Blue: derived using $\tau_b$. Red: derived using $\tau_r$.}
	\label{fig:s_density_cf}
\end{figure}

\begin{figure}
	\centering
	\includegraphics[width=0.6\textwidth]{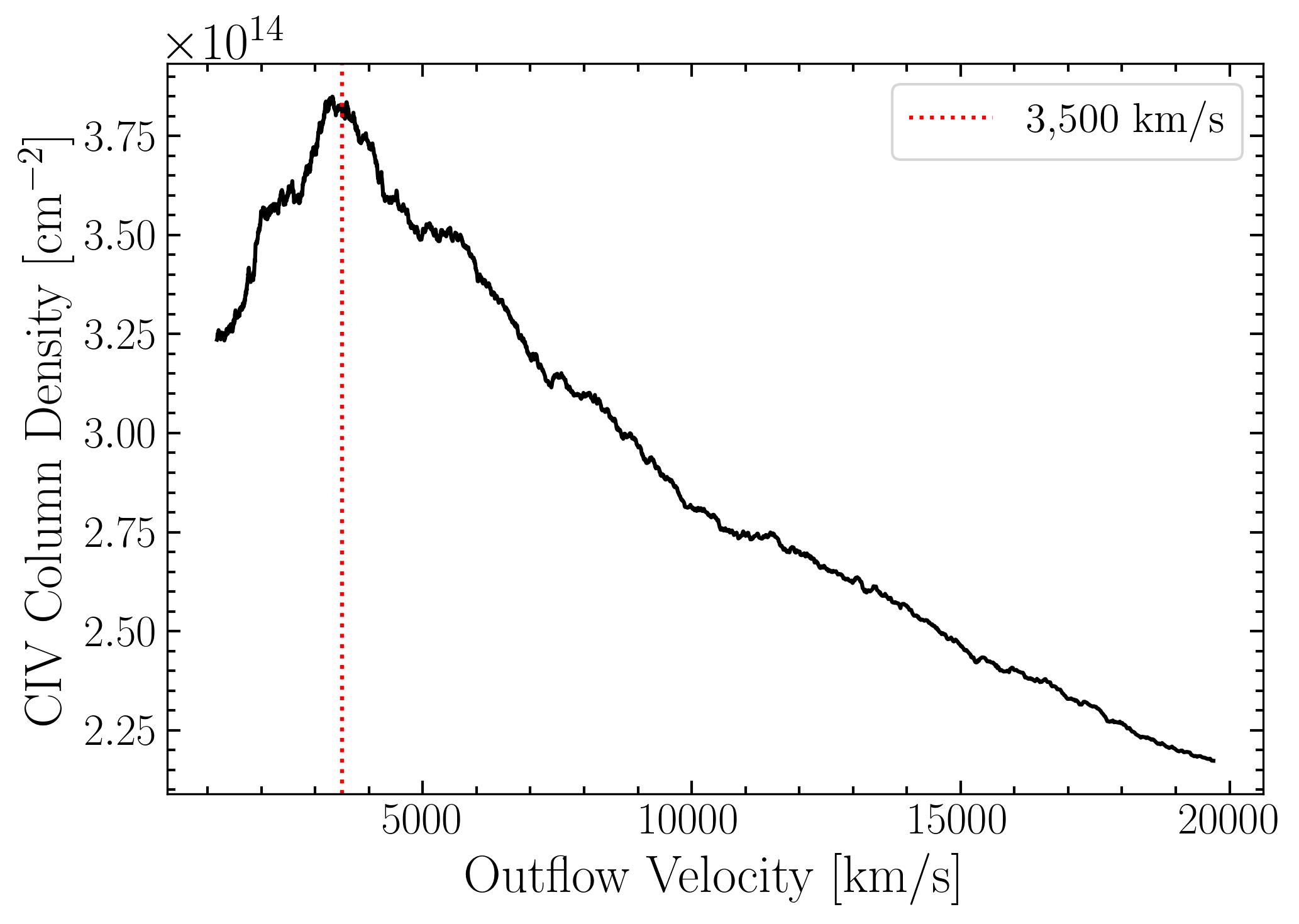}
	\caption{\textbf{\CIV\ column density as a function of outflow velocity.}
	Column density extracted using the sliding-window stacking technique described in Materials and Methods, maintaining a large, fixed number of spectra per stack while smoothly sampling the outflow velocity distribution.}
	\label{fig:s_density_vel}
\end{figure}

\begin{figure}
	\centering
	\includegraphics[width=0.6\textwidth]{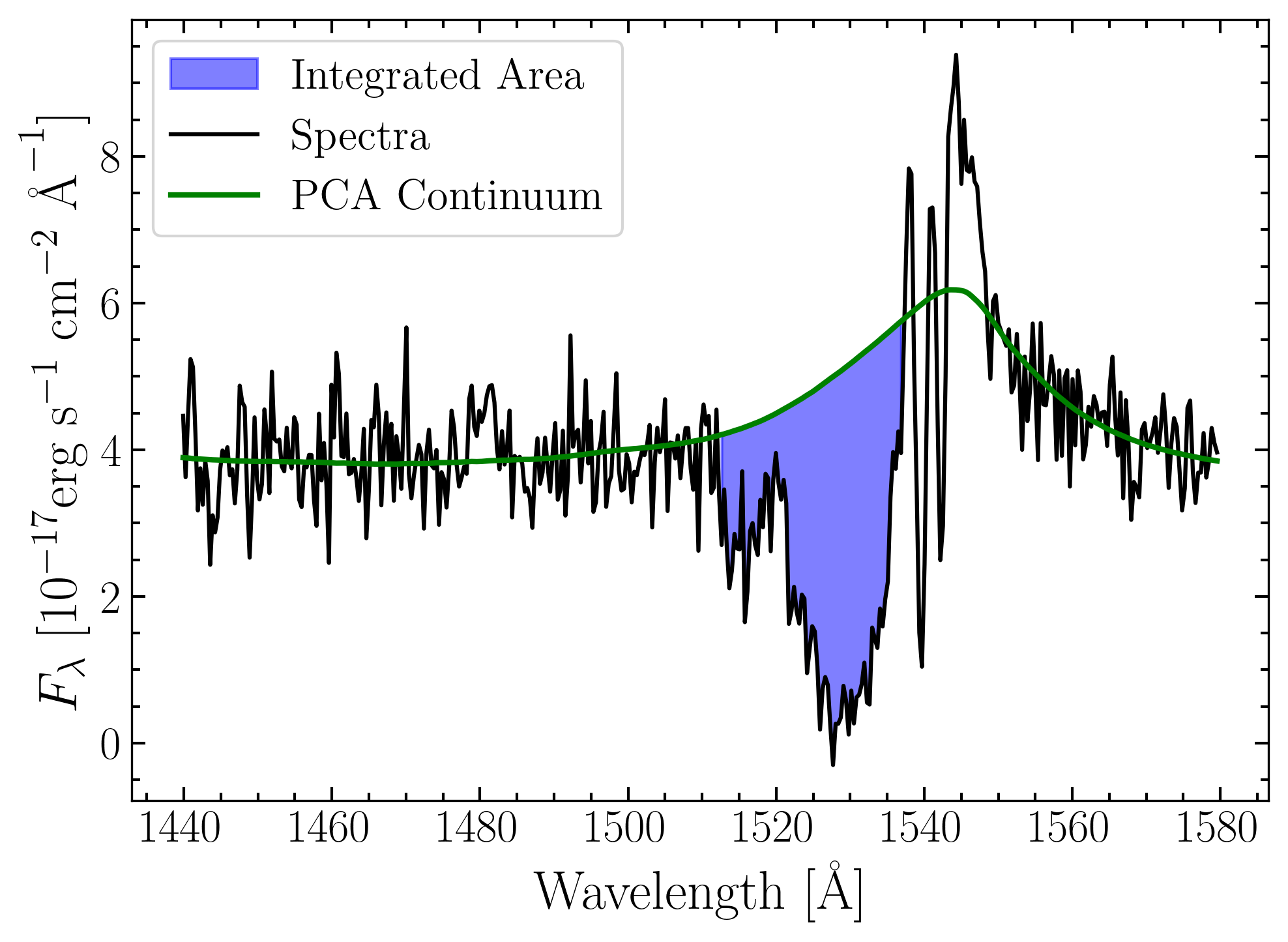}
	\caption{\textbf{Example measurement of absorbed flux for a single trough.}
	Close-up of the \CIV\ region illustrating the measurement of absorbed flux (blue shaded area) by integrating the area between the PCA continuum (green) and the observed spectrum (black) over the wavelengths bounding the trough.}
	\label{fig:s_mdot_example}
\end{figure}

\begin{figure}
	\centering
	\includegraphics[width=\textwidth]{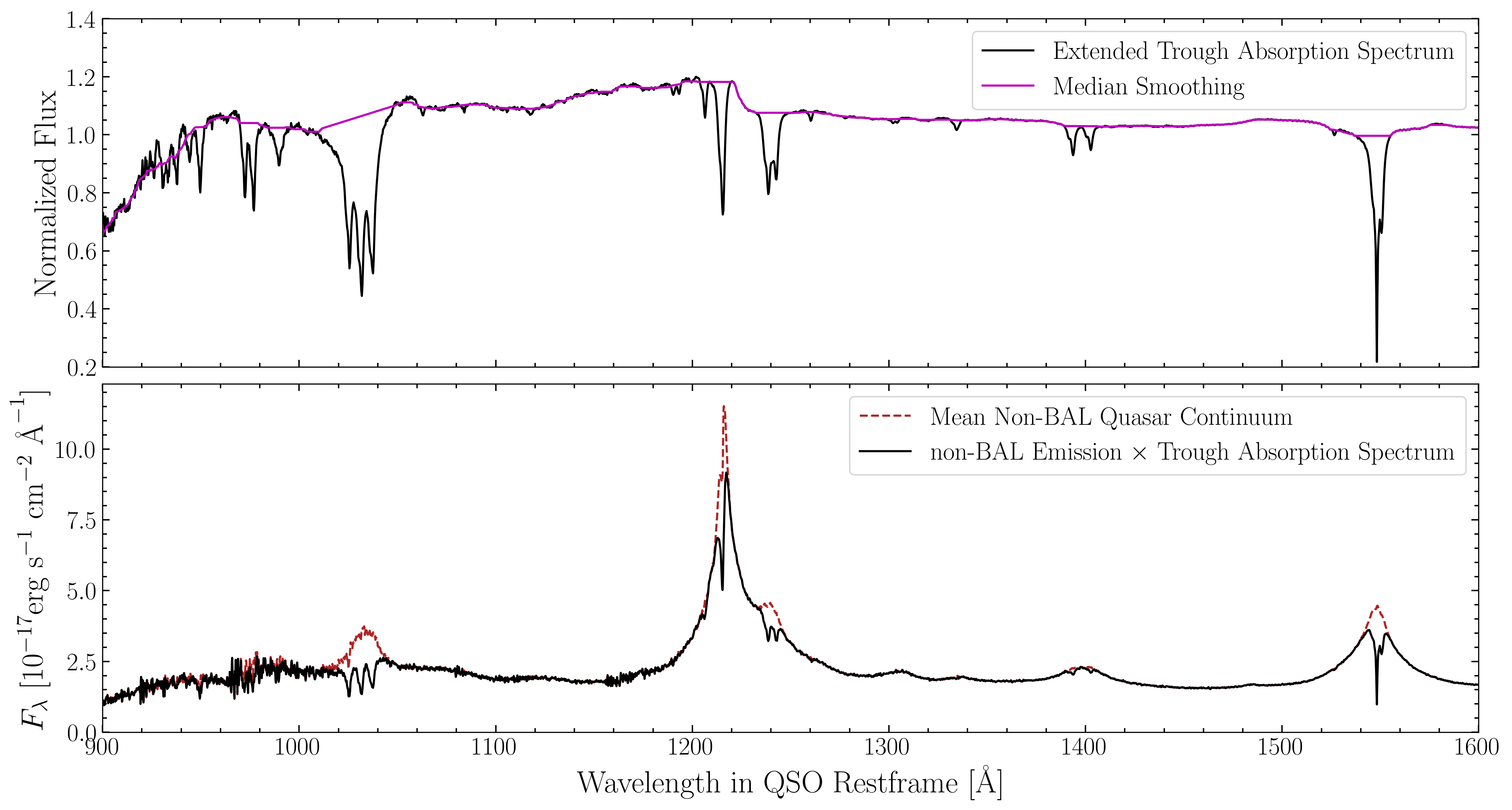}
	\caption{\textbf{Extended composite absorption and emission spectra used to estimate non-\CIV\ absorption.}
	(Top) Mean \CIV\ trough absorption spectrum built using SDSS pipeline continua, extending bluewards to the \NV, Lyman-$\alpha$, and \OVI\ line regions (black), with its smoothed, unabsorbed continuum (purple) used for renormalization. (Bottom) The mean DR16 non-BAL quasar emission spectrum (dotted red) and the product of this emission spectrum and the renormalized trough absorption spectrum, assuming zero outflow velocity offset (black). The difference between the red and black curves gives the average flux absorption at a given outflow velocity, used to estimate the contribution of ions other than \CIV.}
	\label{fig:s_extstack}
\end{figure}

\begin{figure}
	\centering
	\includegraphics[width=0.6\textwidth]{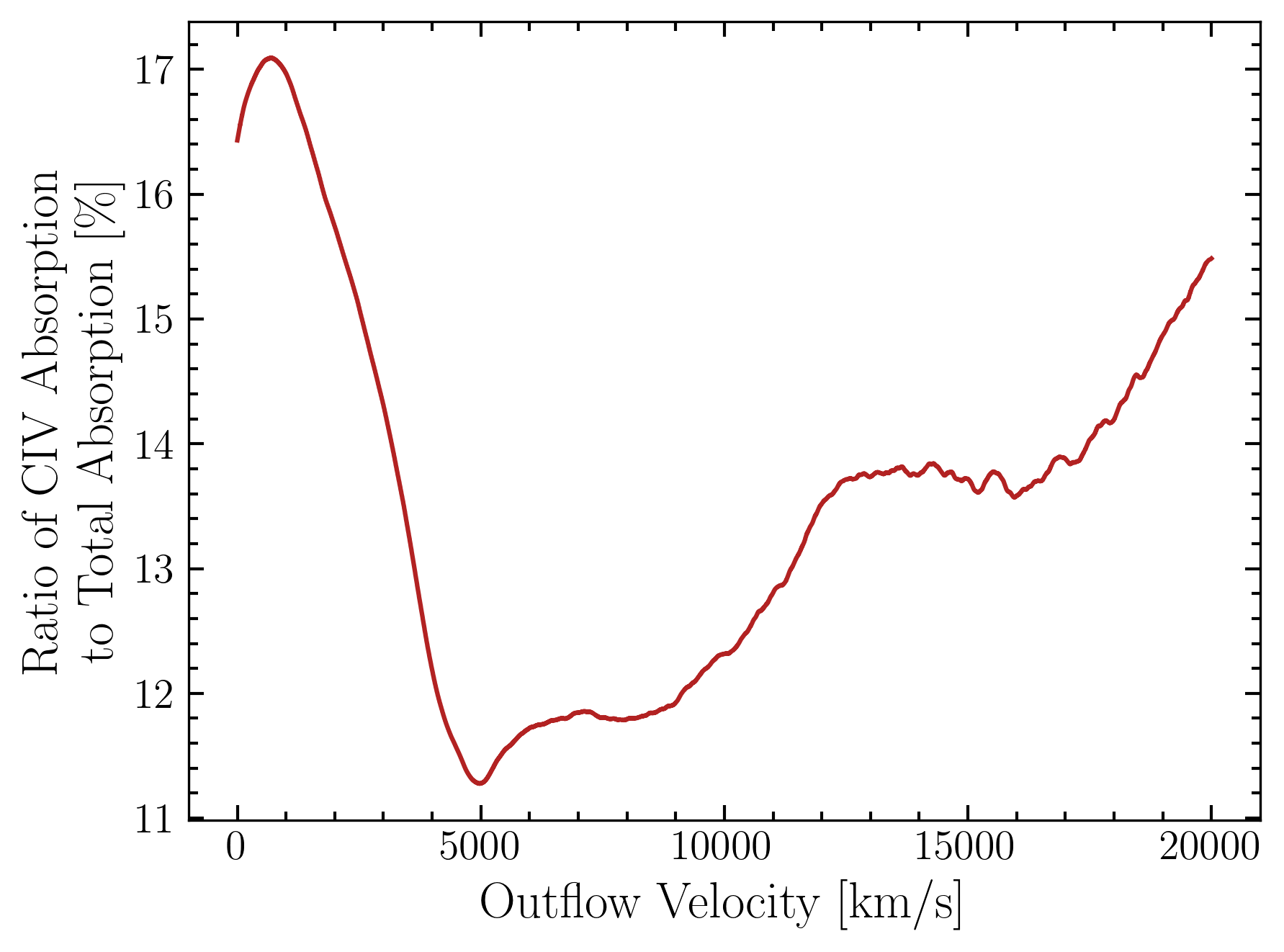}
	\caption{\textbf{Fractional contribution of \CIV\ to total absorption as a function of outflow velocity.}
	Computed from the difference, at the position of the \CIV\ trough, between the red and black curves in Fig.~\ref{fig:s_extstack} (bottom panel), as the two spectra are shifted to different relative velocity offsets.}
	\label{fig:s_civratio}
\end{figure}

\begin{figure}
	\centering
	\includegraphics[width=\textwidth]{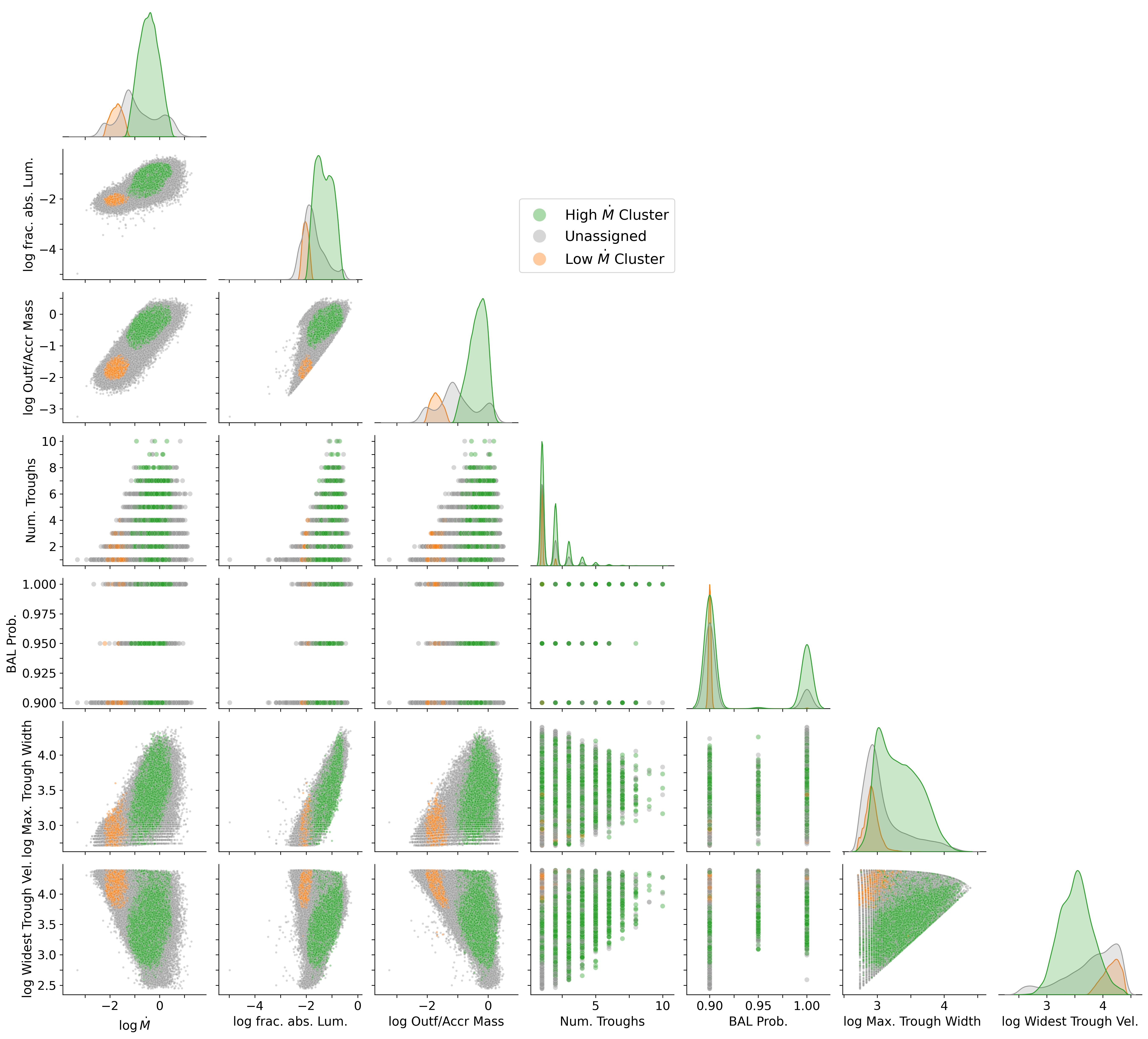}
	\caption{\textbf{Joint distributions for the two outflow populations.}
	Distributions for the low-mass-outflow-rate (orange) and high-mass-outflow-rate (green) populations, and objects not assigned to either by DBSCAN clustering, in the space of outflow mass rate, absorbed luminosity fraction, mass outflow to accretion ratio, number of troughs, BAL probability, maximum trough width, and widest trough velocity.}
	\label{fig:s_populations}
\end{figure}

\begin{figure}
	\centering
	\includegraphics[width=\textwidth]{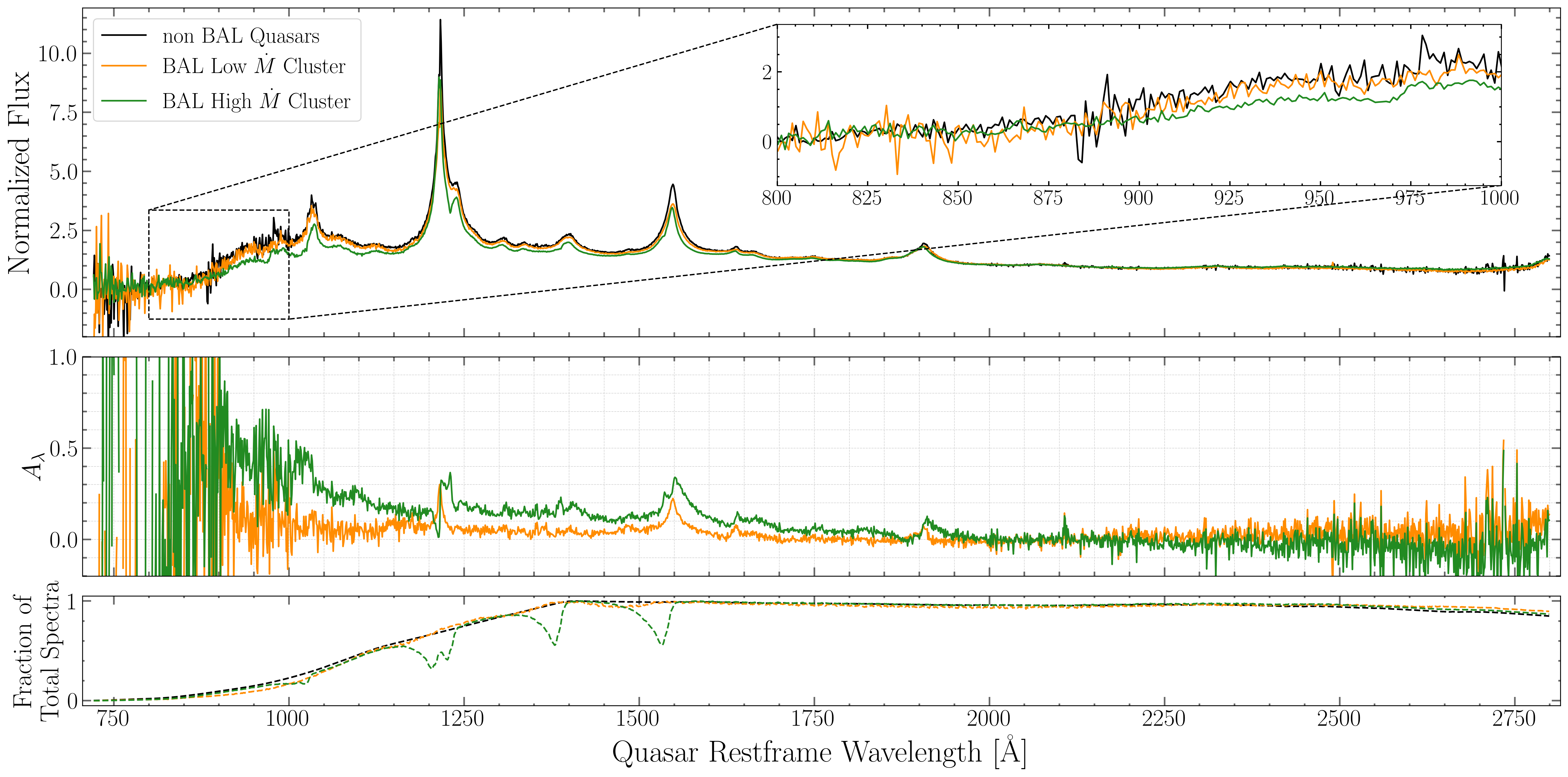}
	\caption{\textbf{Dust reddening of the two outflow populations.}
	(Top) Continuum stacks for non-BAL quasars (black) and the two outflow populations (orange, green). (Middle) Excess extinction inferred from the difference between each outflow population continuum and the non-BAL continuum. (Bottom) Fraction of spectra contributing to each stack as a function of wavelength, illustrating the effect of trough masking.}
	\label{fig:s_dust}
\end{figure}

\begin{figure}
	\centering
	\includegraphics[width=0.6\textwidth]{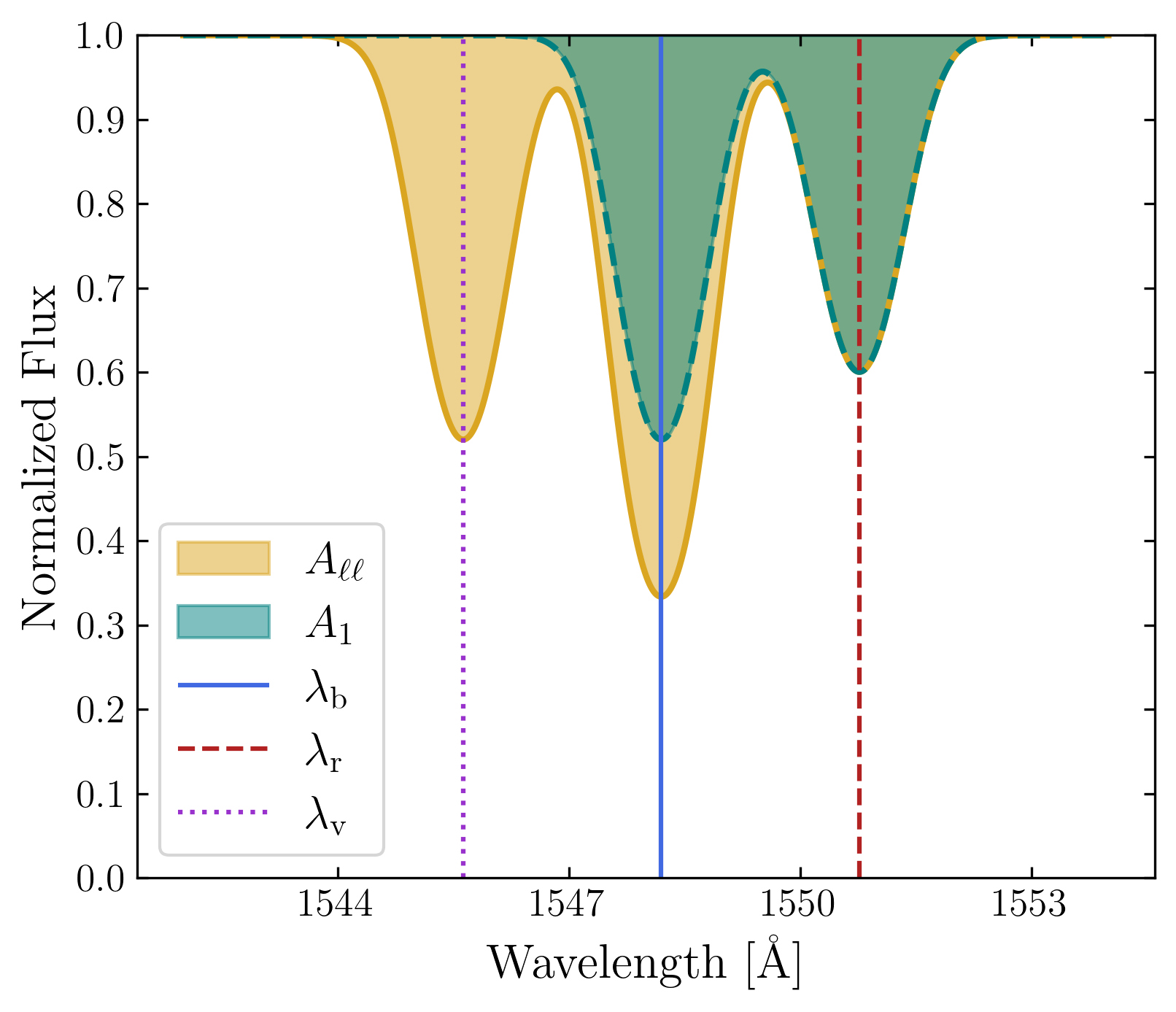}
	\caption{\textbf{Single-cloud versus line-locked \CIV\ absorption profiles used to estimate $F_{\rm masked}$.}
	Normalized \CIV\ absorption profiles for a single-cloud system (Eq.~\ref{eq:I1}, green) and a line-locked two-cloud system (Eq.~\ref{eq:Ill}, gold), using the optical depths in Fig.~\ref{fig:s_opticaldepths} and covering fraction $f_c=0.9$. The integrated absorption areas are $A_1=1.44$ and $A_{\ell\ell}=2.56$, giving $F_{\rm masked}=(A_{\ell\ell}-A_1)/A_1\approx0.2$.}
	\label{fig:s_masking}
\end{figure}

\begin{figure}
	\centering
	\includegraphics[width=\textwidth]{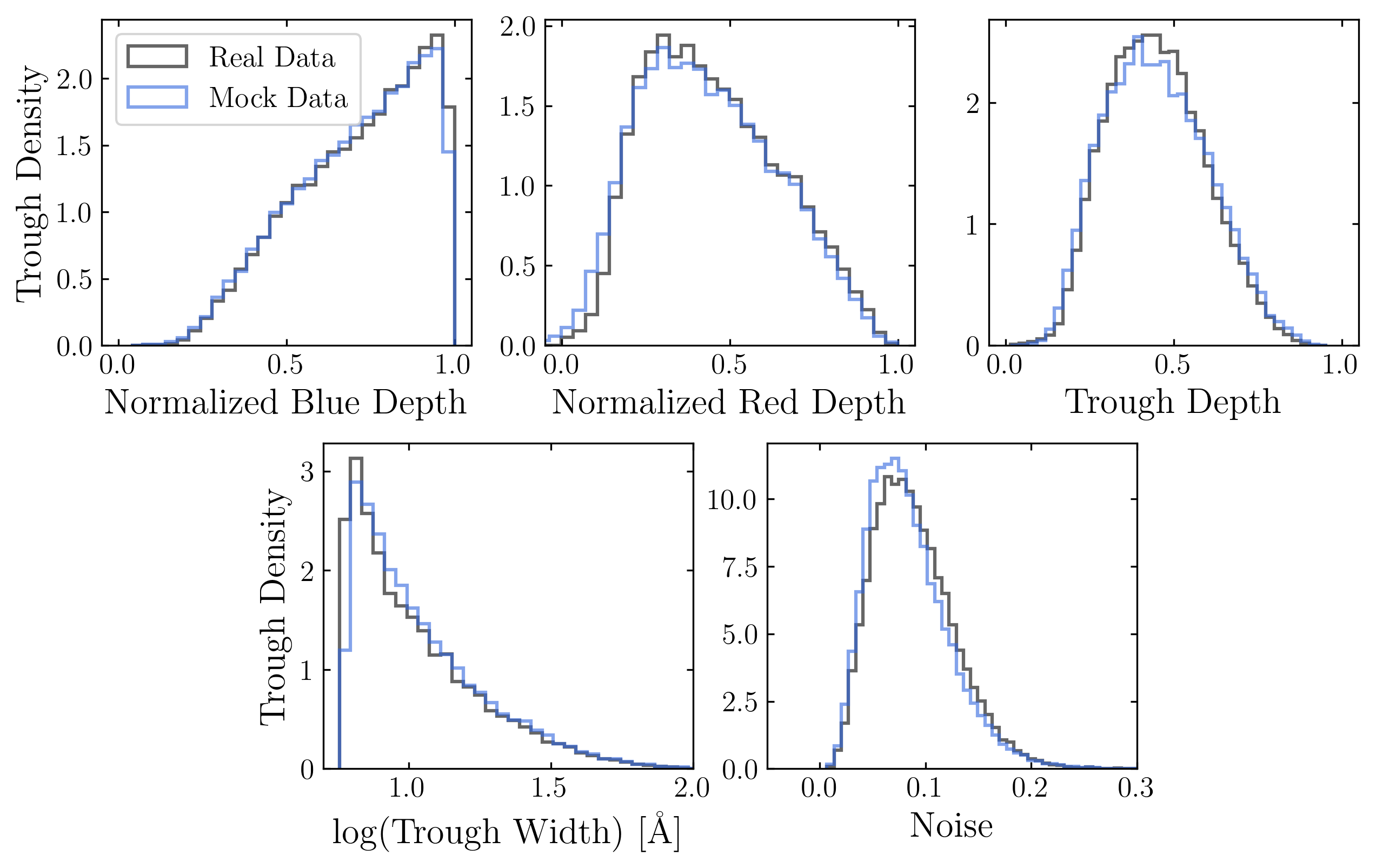}
	\caption{\textbf{Comparison of real and mock trough parameter distributions (two-thirds triplet, one-third doublet mocks).}
	Same comparison as Fig.~\ref{fig:s_paramcomp}, but using the mixed triplet/doublet mock sample found to best match the real data (main text Fig.~1). Overall agreement is comparable to the doublet-only case.}
	\label{fig:s_paramcomp_triplet}
\end{figure}

We tested three trough-edge models against the same KDE sample: a strict inverted boxcar with no slanted modification on either side; a boxcar with the slanted modification always applied on the red side regardless of relative depth; and our fiducial model, in which the modification is applied only when the red (or blue) line depth is shallower than the trough depth (Fig.~\ref{fig:s_troughmodels}). The strict boxcar produces a smearing effect because the trough is deeper than the red line itself, while the always-slanted model produces an artificially pronounced red line. The fiducial model best reproduces the red side of the real stack (Fig.~\ref{fig:s_doubletstack}) and gives the closest match between mock and real trough-parameter distributions. The fraction of spectra on which the stacking algorithm aligns on the blue line, red line, or noise varies modestly across these three models (65/3/32 percent, 77/4/19 percent, and 72/3/25 percent, respectively), but the red-line contamination fraction remains at the few-percent level in every case, supporting the robustness of our contamination estimate to this modeling choice.

\subsubsection*{Status of the low-mass-outflow-rate population}

\cite{Knigge2008} previously identified two classes of BAL outflow based on trough strength, broadly corresponding in our formalism to quasars with BAL probability at or above 0.9. Our low-mass-outflow-rate population corresponds only to the narrow, high-velocity tail of this group, while our high-mass-outflow-rate population spans both of the classes identified by \cite{Knigge2008}. The absence of line locking, minimal dust reddening, and the narrow, high-velocity character of the low-mass-outflow-rate troughs together suggest that this population may not represent genuine outflowing wind material, but rather intervening absorption systems associated with the host galaxy or other foreground structures along the line of sight. We do not include this population in our feedback conclusions, which are based on the line-locking-confirmed, high-mass-outflow-rate population.

\subsubsection*{Limitations}

Our stacking methodology necessarily averages over a diverse population of individual systems with different physical properties and evolutionary states; reported values represent averages over the full sample rather than properties of any single system. High-resolution, high-signal-to-noise observations of individual quasars, which enable direct detection of line locking in specific systems and study of its correlation with other quasar properties \cite{Chen2025}, provide an important complement to our stacked analysis and could help establish whether our averaged measurements are representative of the full population. Our mass outflow rate calculation assumes single scattering in converting absorbed luminosity to outflow momentum; multiple scattering, if present, would increase the inferred momentum transfer efficiency. We measure absorbed flux only over the rest-frame ultraviolet range covered by the BOSS spectrograph and do not account for any additional absorption outside this window. The line-driven outflow we characterize may also represent only one phase of a more complex, multi-phase structure: if the ultraviolet-absorbing gas is the swept-up sheath of a faster, hotter wind driven by a different mechanism \cite{Mas-Ribas2019b}, our measurements would underestimate the total energy injected by the outflow. Our CAK wind model is a simplified, steady-state, single-fluid treatment that neglects non-radial flow, detailed ionization stratification, and any dynamical effects of the accretion disk; while these simplifications could alter the detailed shape of the acceleration profile and the corresponding range of $\epsilon$, the self-consistency of the wide observed line-locking velocity range with $g_\ell/g_{\rm other}\sim1$ (main text) is unlikely to be altered by relaxing them.


\begin{table}
	\centering
	\caption{\textbf{SDSS DR16 \CIV\ trough selection criteria.}
	Quality and selection cuts applied to the initial 159,690 \CIV\ troughs identified by \texttt{baltools} in 397,039 DR16 quasars. All criteria are applied simultaneously.}
	\label{tab:selection}
	\begin{tabular}{lc}
		\\
		\hline
		Selection criterion & Threshold \\
		\hline
		$\chi^2$ from PCA continuum model & $<5$ \\
		$\chi^2$ from SDSS continuum model & $<10$ \\
		Median signal-to-noise ratio (full spectrum) & $>2$ \\
		PCA-normalized flux minimum & $>0.01$ \\
		PCA-normalized flux minimum & $<0.99$ \\
		BAL probability \cite{DR16} & $\ge0.8$ \\
		\hline
		Final number of \CIV\ troughs & 81,388 \\
		\hline
	\end{tabular}
\end{table}

\begin{table}
	\centering
	\caption{\textbf{Comparison of statistical moments between real and mock trough parameter distributions.}
	First four moments (mean, standard deviation, skewness, kurtosis) of the blue depth, red depth, trough depth, trough width, and noise distributions for the real data and for doublet-only mock catalogs, corresponding to Fig.~\ref{fig:s_paramcomp}. Mock values are the mean of 50 independent mock catalog realizations, each comprising 29,332 mock spectra.}
	\label{tab:moments}
	\begin{tabular}{lccccc}
		\\
		\hline
		Statistic & Blue depth & Red depth & Trough depth & Trough width & Noise \\
		\hline
		Real mean      & 0.718  & 0.450  & 0.448  & 1.027 & 0.089 \\
		Mock mean      & 0.713  & 0.435  & 0.448  & 1.047 & 0.085 \\
		Real std.\ dev.& 0.191  & 0.198  & 0.143  & 0.227 & 0.038 \\
		Mock std.\ dev.& 0.192  & 0.204  & 0.152  & 0.227 & 0.039 \\
		Real skewness  & $-$0.520 & 0.330 & 0.251  & 1.141 & 0.751 \\
		Mock skewness  & $-$0.524 & 0.232 & 0.262  & 1.141 & 0.944 \\
		Real kurtosis  & $-$0.673 & $-$0.693 & $-$0.403 & 0.903 & 0.773 \\
		Mock kurtosis  & $-$0.643 & $-$0.644 & $-$0.537 & 0.917 & 0.923 \\
		\hline
	\end{tabular}
\end{table}



\end{document}